\documentclass[extra,mreferee]{gji}

\usepackage[pdftex]{color,graphicx}

\usepackage{amssymb}
\usepackage[fleqn]{amsmath}
\usepackage{float}
\usepackage{subfig}
\usepackage{bm}

\usepackage{times}

\usepackage{color}
\usepackage{booktabs}

\usepackage{multirow}

\usepackage[%
  pdfstartview=FitV,%
  bookmarks=false,%
  bookmarksopen=false,%
  breaklinks=true,%
  colorlinks=true,%
  linkcolor=blue,anchorcolor=blue,%
  citecolor=blue,filecolor=blue,%
  menucolor=blue,%
  urlcolor=blue]{hyperref}
  
\usepackage{rotating}

\newcommand{\alphav}{\bm \alpha}
\newcommand{\betav}{\bm \beta}
\title[Core and lithospheric fields]{Co-estimation of core and lithospheric magnetic fields\\ by a maximum entropy method}

\author[Otzen et al.]{Mikkel Otzen, Christopher C. Finlay and Clemens Kloss\\
Division of Geomagnetism and Geospace, Department of Space Research and Technology\\ Technical University of Denmark, Kgs. Lyngby, Denmark,\\
E-mail: cfinlay@space.dtu.dk \\
}  
\date{Received XXX; in original form XXX}
\pagerange{\pageref{firstpage}--\pageref{lastpage}}
\volume{XXX}
\pubyear{xxxx}
\begin{document}
\maketitle
\vspace{-1.0cm}
\begin{summary}
Satellite observations of the geomagnetic field contain signals generated in Earth's interior by electrical currents in the core and by magnetized  rocks in the lithosphere.  At short wavelengths the lithospheric signal dominates, obscuring the signal from the core. Here we present details of a method to co-estimate separate models for the core and lithospheric fields, which are allowed to overlap in spherical harmonic degree, that makes use of prior information to aid the separation.  Using a maximum entropy method we estimate probabilistic models for the time-dependent core field and the static lithospheric field that satisfy constraints provided by satellite observations 
while being consistent with prior knowledge of the spatial covariance and expected magnitude of each field at its source surface. 

For the core field, we find that between spherical harmonic degree 13 and 22 power adds coherently to the established structures, and present a synthetic test that illustrates the aspects of the small scale core field that can reliably be retrieved.  For the large scale lithospheric field we also find encouraging results, with the strongest signatures below spherical harmonic degree 13 occurring at locations of known prominent lithospheric field anomalies in north-Eastern Europe, Australia and eastern North America.   Although the amplitudes of the small scale core field and large scale lithospheric field are underestimated we find no evidence that obvious artefacts are introduced.  Compared with conventional maps of the core-mantle boundary field our results suggest more localized normal flux concentrations close to the tangent cylinder, and that low latitude flux concentrations occur in pairs of opposite polarity.  Future improvements in the recovery of the small scale core field and large scale lithospheric field will depend on whether more detailed prior information can be reliably extracted from core dynamo and lithospheric magnetisation simulations.

\end{summary}
%===============================================================================
%===============================================================================

\section{Introduction}
\label{intro}

Earth's magnetic field is a result of sources located both within the Earth and above its surface in the upper atmosphere \citep{Chapman_1940,Langel_1998,Olsen_Stolle_2012}.  Spherical harmonic analysis indicates that internal sources are responsible for the majority of the field \citep{Gauss_1839, Hulot_2015}.  The spatial power spectrum of the internal field at Earth's surface is steep at low degree (up to approximately spherical harmonic degree 13) and essentially flat at higher degree \citep{Lowes:1974}, indicating two sources: one deep within the planet and one located near to the surface \citep{Voorhies_etal_2002,Voorhies_2004}. These deep and shallow sources are thought to correspond to the core dynamo and lithospheric magnetisation. 

Measurements of Earth's magnetic field from space have provided an increasingly detailed picture of the magnetic field due to internal sources.  The MAGSAT mission \citep{Langel_1982} delivered the first set of vector measurements with global coverage allowing the change in the slope of the spatial power spectra, between the wavelengths where core and respectively lithospheric sources dominate, to be definitively observed \citep{Langel_Estes_1982}.  With more recent satellite missions, in particular the CHAMP \citep{Reigber_2002} and \textit{Swarm} \citep{Friis_2006,Olsen_2018} missions, it is possible to determine the internal field spectrum out to 
beyond degree 130 \citep{Maus_2010,Olsen_2017}.  Knowledge of the small scale core field (which we define here as above spherical harmonic degree 13) has on the other hand advanced rather little since the time of MAGSAT due to it being obscured by the lithospheric field. The conventional approach is to estimate a single internal spherical harmonic field model and then to truncate at degree 13 to study the core field, for example when plotting maps of the radial field at the core-mantle boundary (CMB) \citep[e.g.][]{Cain_1989, Olsen_etal_2014, Sabaka_2020}.

Truncation at a fixed spherical harmonic degree has however limitations when seeking to isolate the core field.  Abrupt truncation in spectral space may cause ringing in physical space \citep{Whaler:1981,Gubbins_2007}.  Furthermore, the lithospheric field does, of course, not stop at degree 14, and there will be some contribution to the internal field below 13 from lithospheric sources.   Most seriously all information on the small scale core field above degree 13 is lost.  

In the 1980s it was suggested that a better way to estimate the core field would be to minimize suitable norms of the field complexity at the CMB \citep{Shure_1985,Gubbins_1985}.  This approach, known as spatial regularization of the field, has been widely adopted for studying the core field over historical \citep{Bloxham_Gubbins_Jackson_1989,Jackson_2000} and paleomagnetic timescales \citep{Korte_2011,Panovska_etal_2018} when data coverage is sparse; it makes use of prior information from seismology on the depth of the CMB, along with asking for a field that is simple in a specific way (as measured by a chosen regularization norm) at the source radius.  A drawback is that traditional regularizations norms, such as the squared value of the radial field or the horizontal gradient of the radial field integrated over CMB, or Ohmic heating norms \citep{Gubbins_1976,Jackson_2000}, strongly penalize small length scales and typically cause the spatial power spectrum to decay in an unphysical fashion above degree 13 \citep{Backus_1988,Buffett_Christensen_2007}. 

Geodynamo simulations for which the magnetic Reynolds number is of order 1000, as expected in Earth's core \citep{Christensen_2004,Lhuillier_2011}, involve localized, high amplitude, flux features and spatial spectra at the CMB that are rather flat, decreasing only very slowly at spherical harmonic degrees 10 to 30 \citep[see, for example,][]{schaeffer2017turbulent,aubert2017spherical,Sheyko_2018}. \citet{Jackson_2003} and \citet{Jackson_etal_2007} showed that regularization norms based on the entropy of the radial field at the CMB allowed the estimation of core fields with flatter spatial spectra and localized, high amplitude flux features. The entropy regularization technique was adapted to time-dependent spherical harmonic field models by \citet{Gillet_etal_2007} and applied to satellite observations from the {\O}rsted, SAC-C and CHAMP missions by \citet{Finlay2012}.  A drawback in these studies was the need to abritrarily pick a value for the so-called default parameter (the magnitude of the radial field expected in the absence of data constraints) that controlled the width of the entropy distribution and hence the sharpness of the field structures \citep{Maisinger_etal_2004,Jackson_etal_2007}.  \citet{Jackson_2003} and \citet{Jackson_etal_2007} focused on default values around 10 $\mu$T for the core field, while \citet{Gillet_etal_2007} used 30 $\mu$T.  Despite many desirable features, entropy-based field reconstruction techniques have been little exploited in subsequent years in part due to doubts as to how to pick the troublesome default parameter.

An important conceptual step forward in co-estimating core and lithospheric field sources was made by \citet{Holschneider_etal_2016}. They suggested how various field sources (including the core and lithospheric fields) could be co-estimated within a Bayesian framework making use of prior information, for example on the expected source depth and its correlation (or covariance) structure.   This approach has been used to develop temporal sequences of field models using a Kalman filter algorithm, being applied to the modelling of satellite and ground magnetic observations by \citet{Ropp_2020}, \citet{Ropp_Lesur_2023} and   \citet{Baerenzung_2020,Baerenzung_2022}.  Using a simple correlation function and treating the source depth as a free parameter  \citet{Baerenzung_2020} were able to construct stable maps of the field at the CMB up to spherical harmonic degree 20, although it was found to be difficult to reliably separate the large scale lithospheric field.  More detailed prior information on the covariance between spherical harmonic coefficients in dynamo simulations has also been used in combination with observation-based internal field models up to degree 13 to infer the core field up to degree 30 \citep{Aubert_2015,Aubert_2020}.  On the theoretical side \citet{Baratchart_Gerhards_2017} have shown that core and lithospheric fields can be formally separated if the lithospheric field sources are localized to a sub-region of the spherical surface.  Non-Gaussian field distributions thus seem to aid the separation of fields from different sources, as is well known in other contexts such as independent component analysis \citep[e.g.][]{Hyvarinen_Oja_2000}.

Here we build on the above studies and seek to estimate separate models for the core and lithospheric fields within a Bayesian framework using a maximum entropy method that accounts for spatial covariances found in first principles simulations of the core dynamo and the lithospheric magnetisation.  Similar maximum entropy based techniques have previously been applied to signal separation problems in cosmology \citep{Hobson_2010}.  Section \ref{sec:method} sets out details of our Bayesian model estimation scheme and specifies the prior information used.  Section \ref{sec:data} describes the satellite and ground magnetic observations employed.  Section \ref{sec:results} presents our results, with Appendix \ref{sec:AppA} collecting findings from a synthetic test based on a similar data and modelling setup.  We conclude in Section \ref{sec:Disc_conc} with a discussion of what has been achieved and suggestions for future improvements of the method.

%%%%%%%%%%%%%

\section{Methodology}
\label{sec:method}

\subsection{Geomagnetic field model}
We model Earth's magnetic field $\mathbf{B}$ as a potential field, representing it by the gradient of a scalar potential $V$ such that
\begin{equation}
\label{eq:int_SH}
    \mathbf{B}= - \nabla V \qquad \mbox{where} \quad V=V^{int} + V^{ext},
\end{equation}
with $V^{int}$ the potential due to internal sources and  $V^{ext}$ that due to external sources. Both core and lithospheric sources contribute to $V^{int}$, we represent each by a separate spherical harmonic expansion
\begin{eqnarray}
\label{eq:int_SH}
    V^{int} (r,\theta,\phi) &=& a \sum^{N_C}_{n=1} \bigg(\frac{a}{r}\bigg)^{n+1} \sum^n_{m=0} \Big[g^C_{n,m}(t) \cos m \phi + h^C_{n,m}(t) \sin m \phi\Big] P^{m}_{n}(\cos\theta)\\
    &+&
     a \sum^{N_L}_{n=1} \bigg(\frac{a}{r}\bigg)^{n+1} \sum^n_{m=0} \Big[g^L_{n,m} \cos m \phi + h^L_{n,m} \sin m \phi\Big] P^{m}_{n}(\cos\theta)
\end{eqnarray}
where $(r, \theta, \phi)$ are geocentric spherical polar coordinates, $a$ is the Earth's mean spherical reference radius, $n$ is the degree of the spherical harmonic, $m$ the order of the spherical harmonic and $P^{m}_{n}(\cos\theta)$ are associated Legendre functions. $g^L_{n,m}$ and $h^L_{n,m}$ are spherical harmonic coefficients describing the lithospheric field, assumed here to be static, and here considered up to a maximum degree of $N_L=120$.   $g^C_{n,m}(t)$ and $h^C_{n,m}(t)$ are time-dependent spherical harmonic coefficients for the core field, considered up to a maximum degree $N_C=30$.  These are expanded in time using a B-spline basis, of order 6 and with a 0.5 year knot spacing,
\begin{equation}
    g^C_{n,m}(t) = \sum_{k}g^{C}_{n,m,k}\mathcal{B}_{k}(t). 
    \label{eq:core_lith_sep}
\end{equation}
where $\mathcal{B}_{k}$ is the kth basis function of the order 6 B-splines. We collect the coefficients describing the core field in a vector $\mathbf{m}^C = \left\{ g^{C}_{n,m,k}, h^{C}_{n,m,k} \right\}$ and the coefficients describing the lithospheric field in a vector $\mathbf{m}^L = \left\{ g^{L}_{n,m}, h^{L}_{n,m} \right\}$.  Note that the core and lithospheric field representations overlap between spherical harmonic degrees 1 and 30, additional prior information is therefore needed in order to separate them.   

As in the CHAOS-7 model \citep{Finlay_CHAOS7:2020} these internal field coefficients are supplemented by model coefficients $\mathbf{m}^{ext}$
 describing the external field, and coefficients $\mathbf{m}^q$ describing the in-flight alignment of the vector magnetometers on each satellite, to give the full model vector $\mathbf{m}=\left[\mathbf{m}^C, \mathbf{m}^L,\mathbf{m}^{ext},\mathbf{m}^q \right]^T$.
 
\subsection{Bayesian model estimation} 
\subsubsection{Entropic priors for the core and lithospheric fields}
We make use of prior information regarding the radial component of the core and lithospheric fields at their respective source surfaces, the CMB and Earth's surface.  The radial field is evaluated at each source surface on an approximately equal area grid, and values are collected into vectors $\mathbf{b}^{C}$ and $\mathbf{b}^{L}$ for the core and lithospheric fields respectively. Such knowledge of the radial field at the source surface completely defines the potential due to an internal source. These are related to the spherical harmonic model coefficients discussed in the previous section by
\begin{equation}
    \mathbf{b}^{C}(t_p) = \mathbf{G}^{C,t_p}\mathbf{m}^C  \quad \mbox{and} \quad \mathbf{b}^{L} = \mathbf{G}^L\mathbf{m}^L 
\end{equation}
with $\mathbf{G}^{C,t_p}$ and $\mathbf{G}^L$ being matrices that synthesize the radial field from the relevant spherical harmonic model coefficients, for the core field at some epoch $t_p$.

Knowledge regarding the spatial covariance of each field at its source surface is provided in the form of a-priori model covariance matrices $\mathbf{C}_C$ and $\mathbf{C}_L$, with lower triangular Cholesky factors $\mathbf{L}_C$ and $\mathbf{L}_L$ that can be used to transform $\mathbf{b}^{C}$ and $\mathbf{b}^{L}$ to latent variables $\mathbf{x}^{C}$ and $\mathbf{x}^{L}$ such that  
\begin{equation}
    \mathbf{b}^{C}(t_p)  = \mathbf{L}_{C}\mathbf{x}^{C}(t_p) \quad \mbox{and} \quad \mathbf{b}^{L}  = \mathbf{L}_L\mathbf{x}^L. 
    \label{Eqn:latent}
\end{equation}
 The latent variables $\mathbf{x}^{C}$ and $\mathbf{x}^{L}$ therefore describe the core and lithospheric radial fields at their source surfaces in a space where their elements are normalized and decorrelated, as is appropriate for the application of maximum entropy methods \citep{Maisinger_etal_2004}.  
 
 It is assumed that $\mathbf{x}^{C}$ and $\mathbf{x}^{L}$ are each described by an entropic probability density function
\begin{equation}
    P(\mathbf{x}^{C}) \propto \exp\left[ \lambda_C S (\mathbf{x}^{C})\right]
     \quad \mbox{and} \quad P(\mathbf{x}^{L}) \propto \exp\left[ \lambda_L S (\mathbf{x}^{L}) \right], 
\end{equation}
where $S$ is the information entropy for variables that can take both positive and negative values \citep{Gull_Skilling_1990,Hobson_Lasenby_1998}
\begin{equation}
    \label{eq:entropy_pos}
    S[\mathbf{x},\omega] = \sum\limits_{i=1}^{M} \left[ \psi_i - 2\omega - x_i \log\left(\frac{\psi_i + x_i}{2\omega}\right) \right]
\end{equation}
where $M$ is in our case the number of grid points on the spherical surface, $\psi_i = \sqrt{x_i^2 + 4\omega^2}$, and $\omega$ is a so-called 'default' parameter that defines the width of the entropy function.  The information entropy function $S$ is a measure of the amount of uncertainty inherent in the distribution of values $\mathbf{x}$ \citep{Shannon_1948,Jaynes_2003}, the form we use follows from requirements of subset independence, coordinate invariance, system independence and scaling \citep{Skilling_1988}. In the geomagnetic context it can be thought of as measuring the number of ways a given distribution of radial field on the source surface can be arranged from elementary flux bundles \citep{Jackson_2003,Jackson_etal_2007}; fields with larger entropy are simpler in the sense that they can be arranged in more ways. 

Assignment of an entropic prior is argued to be an appropriate choice in the absence of precise information as to the form of a prior pdf \citep{Skilling_1989,Hobson_etal_1998}, and it is compatible with possibly non-Gaussian distributions of $\mathbf{x}$.  Maximizing the entropy essentially broadens the distribution of  $\mathbf{x}$ as much as possible without violating the available observational constraints.  The resulting distribution therefore agrees with what is known but expresses maximum uncertainty with respect to all other matters \citep{Jaynes_1968}. Maximizing the entropy does not introduce additional correlations amongst the latent variables \citep{Gull_Skilling_1984}.

The factors $\lambda_C$ and $\lambda_L$ appearing in the entropic pdfs are scaling factors. We are able to set these equal to 1 because $\mathbf{x}^{C}$ and $\mathbf{x}^{L}$ have already been normalized via the transform of $\mathbf{b}^{C}$ and $\mathbf{b}^{L}$  to the latent space \citep{Hobson_2010}.  The transform to latent space using the a-priori covariance functions also ensures the entropy is computed from uncorrelated variables, an important condition for the maximum entropy method. 

To put into practice the above scheme we require prior information concerning the covariance structure of the radial fields on the source surfaces and the expected widths of the distributions of $\mathbf{x}$ for each source.  We obtain these from first principles simulations of the core dynamo and the lithospheric magnetisation. Full details are provided in \citet{Otzen_2022} only a short summary is given here.

For the core field prior, we use an ensemble of realizations of the core field produced by versions of the coupled-Earth dynamo of \cite{Aubert:2013}. This numerical dynamo is known for producing field structures and patterns of secular variation similar to those observed over the past centuries.  To start with we used a collection of radial fields realizations, well separated in time, generated by the original version of the coupled-Earth dynamo \citep{Aubert:2013} that has been used in previous field modelling and data assimilation studies \citep{barrois2017contributions, Ropp_2020}.  To this we added radial field realizations from a long run of an updated version (71\% of path) of the coupled-Earth dynamo \citep{Aubert:2021}. Although these two cases involve different control parameters they lie on a path through control parameter space along which the field morphology is essentially invariant \citep{aubert2017spherical}.  We finally augmented our set of realizations by carrying out rotations of the simulated core fields by an arbitrary amount in longitude, this was possible because the covariance functions we use do not depend on longitude and this enabled us to work with a larger ensemble.  In all this resulted in an ensemble of 5688 core field realizations up to spherical harmonic degree 30. To be more consistent with the observed field we also adjusted the dipole fields from the dynamo simulations replacing the $n=1$ coefficients with random samples from normal distributions with mean values of $\bar{g}^C_{1,0}=-29000$\,nT, $\bar{g}^C_{1,1} = 0$\,nT, and $\bar{h}^{C}_{1,1}=0$\,nT and standard deviations of $5000$nT, $3000$nT, and $3000$nT respectively, the latter being similar to those seen in the dynamo realizations. The spread of the power spectra from the resulting ensemble encompasses the observed internal field \citep[e.g.][]{Finlay_CHAOS7:2020} up to spherical harmonic degree 13 \citep{Otzen_2022}. 

\begin{figure}
(a) \qquad \qquad \qquad \qquad \qquad \qquad \qquad \qquad \qquad \qquad \qquad  (b)\\
\includegraphics[width=0.5\textwidth]{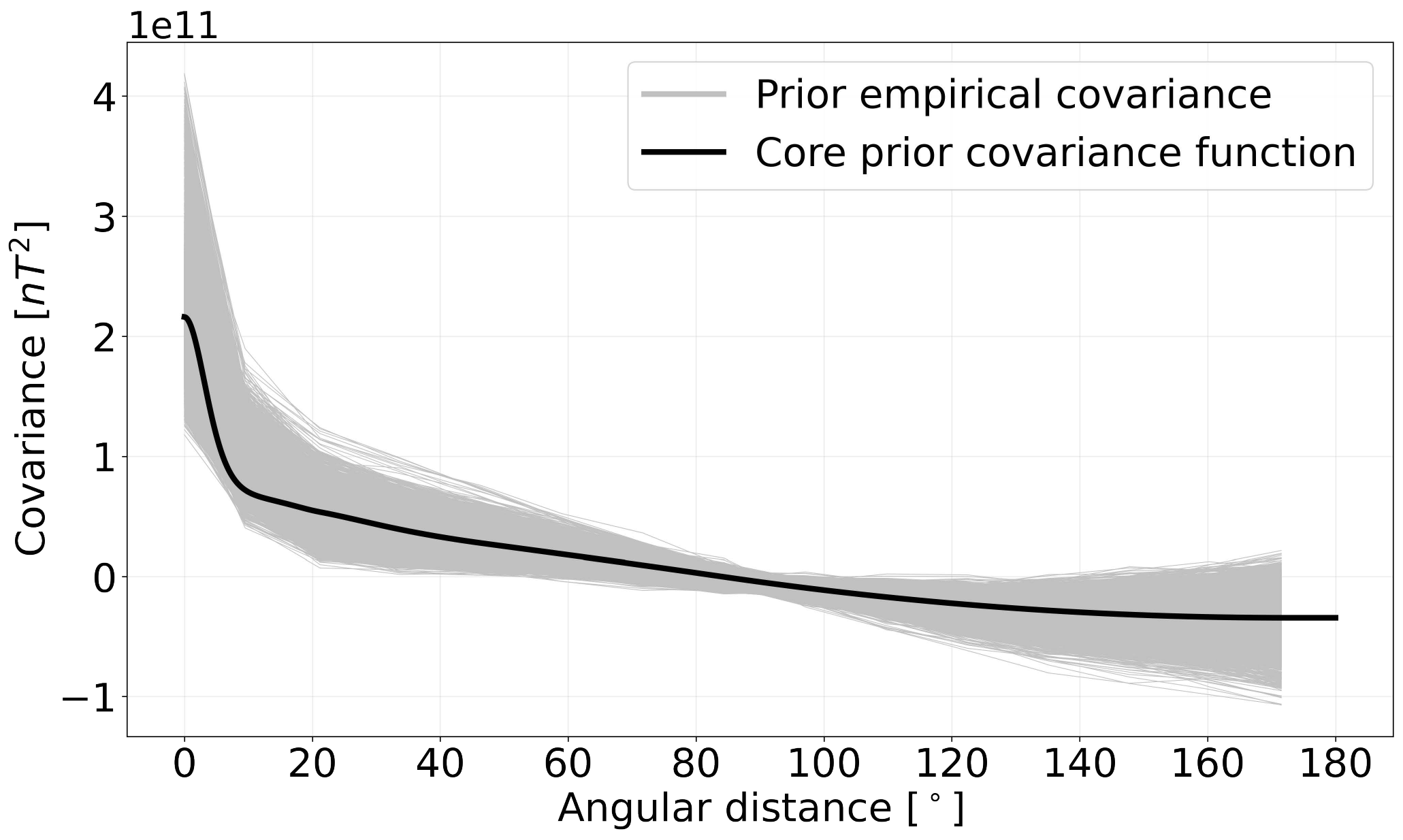}\includegraphics[width=0.5\textwidth]{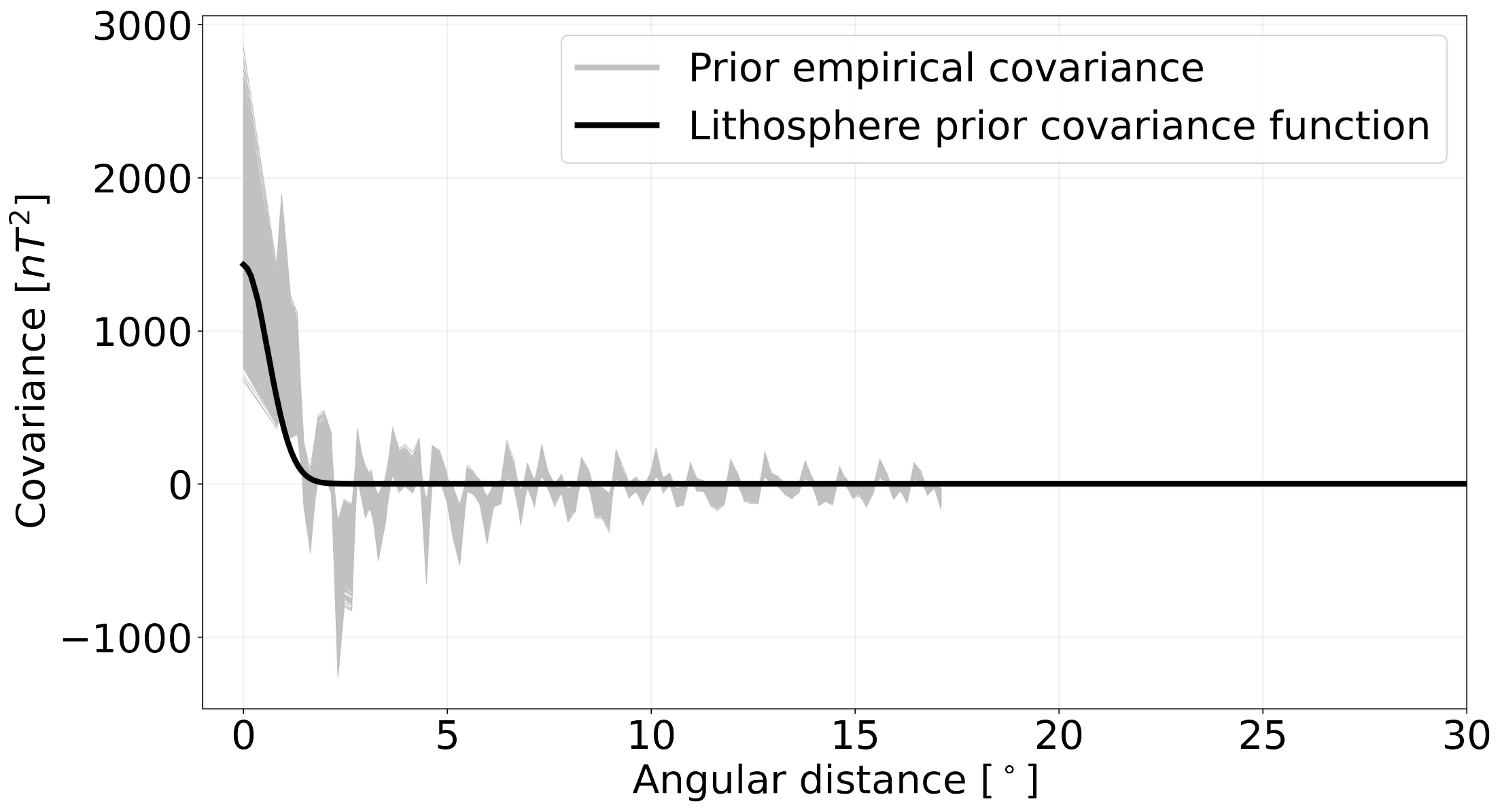}
(c) \qquad \qquad \qquad \qquad \qquad \qquad \qquad \qquad \qquad \qquad \qquad  (d)\\
\includegraphics[width=0.5\textwidth]{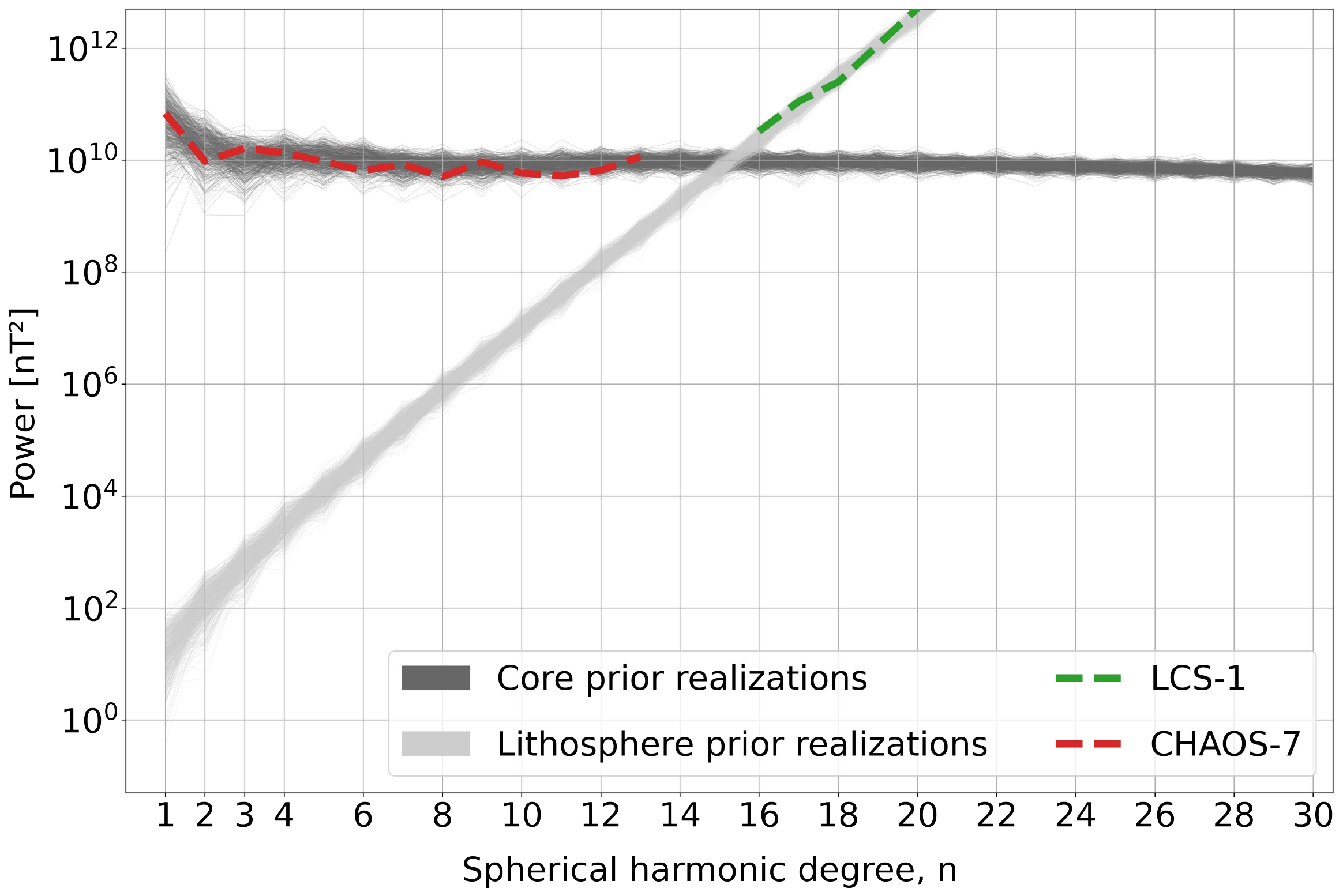}\includegraphics[width=0.5\textwidth]{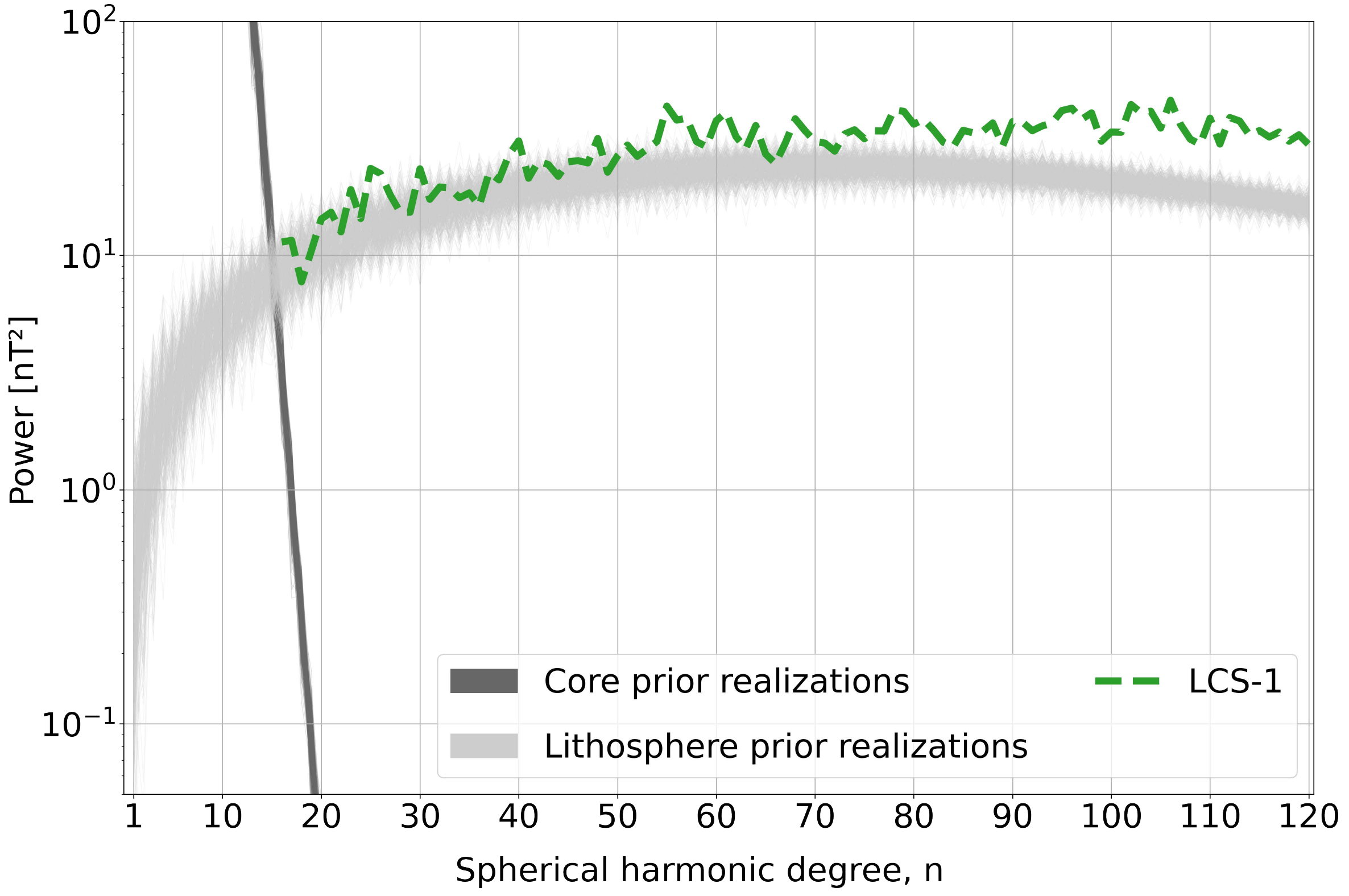}
(e) \qquad \qquad \qquad \qquad \qquad \qquad \qquad \qquad \qquad \qquad \qquad  (f)\\
\includegraphics[width=0.5\textwidth]{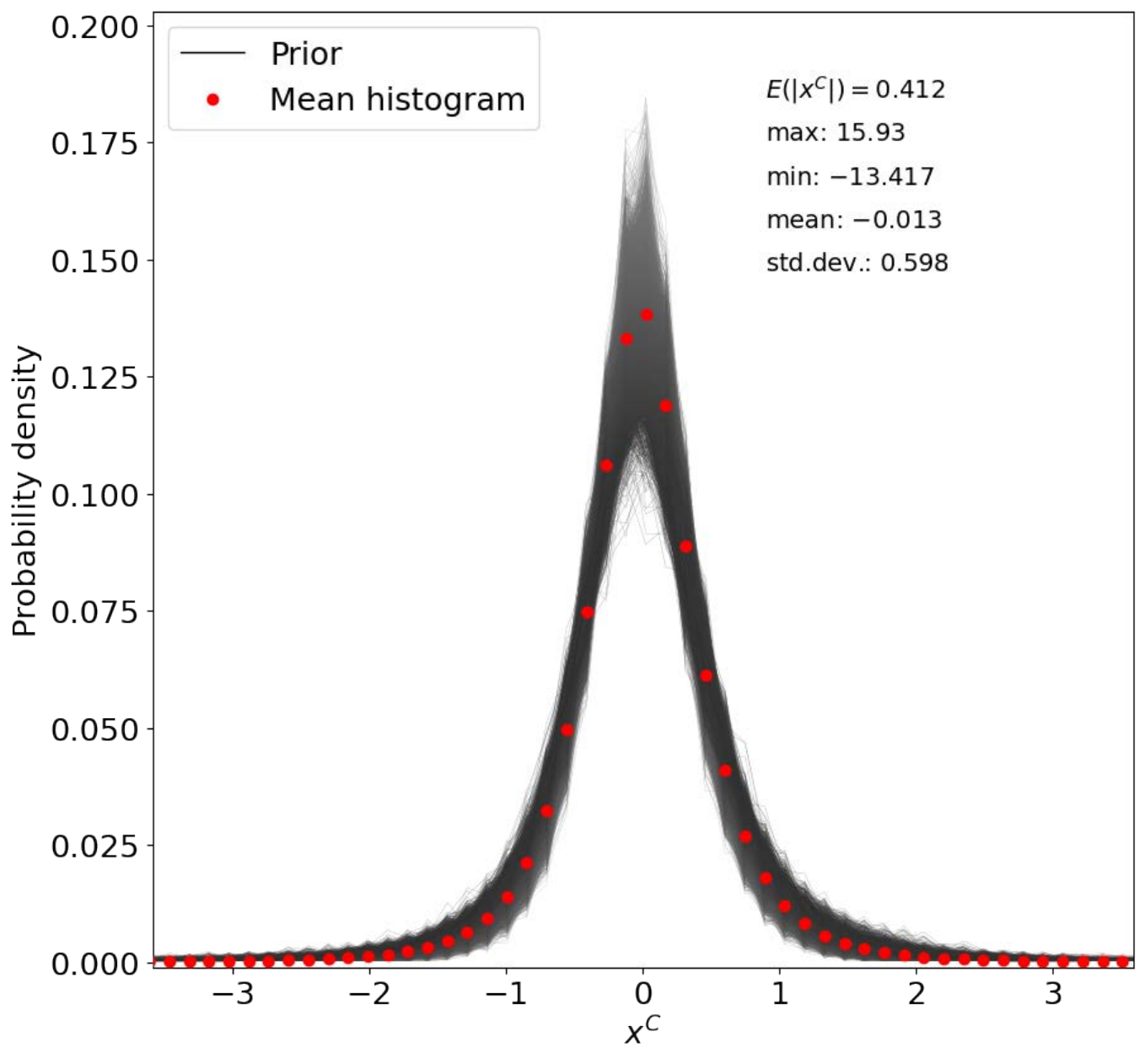}\includegraphics[width=0.5\textwidth]{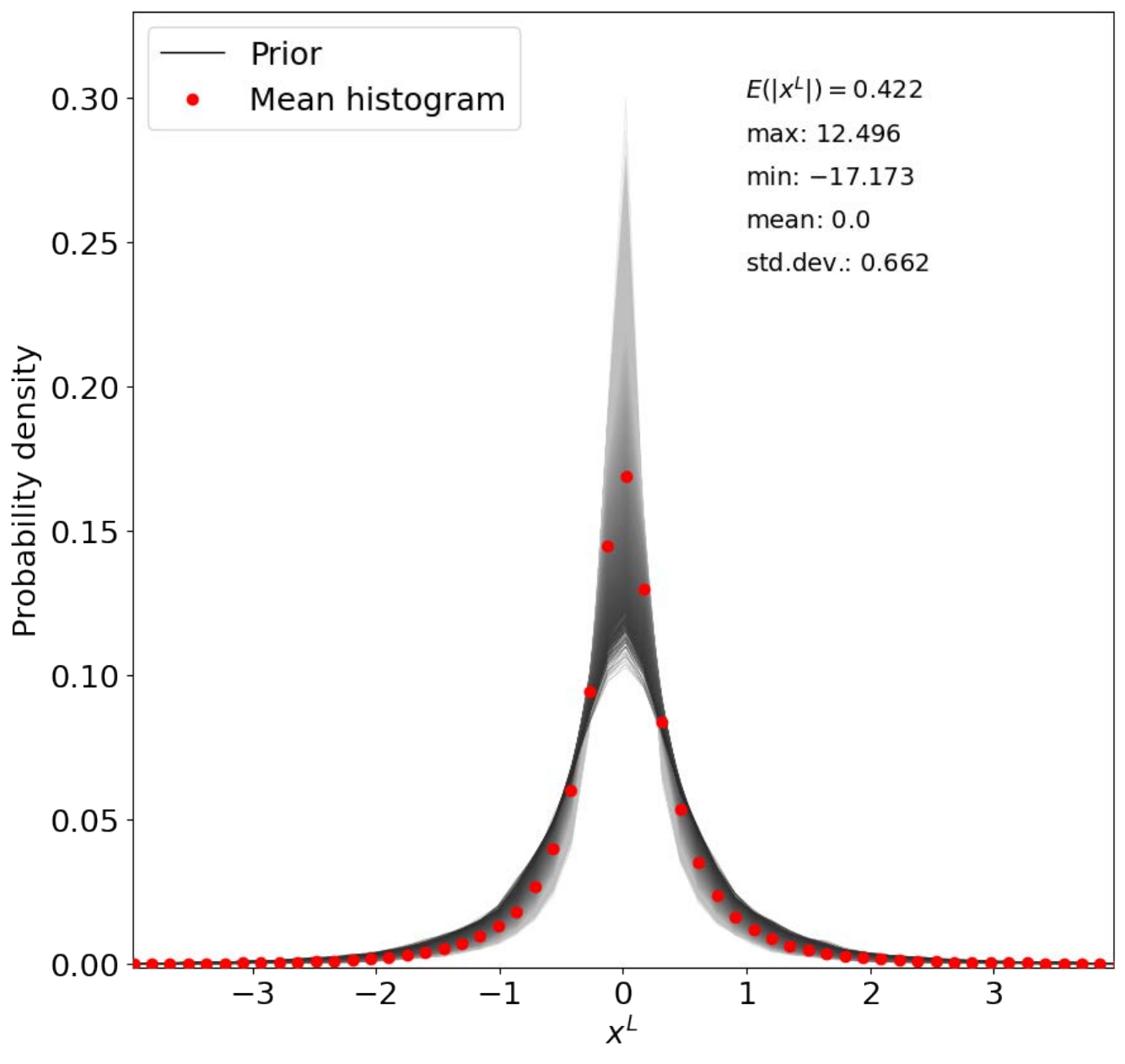}
 
 \caption{Statistical properties of the prior ensemble of core and lithospheric fields.  Top row: Empirical covariance functions (a) for the CMB radial field, (b) for the lithospheric radial field at Earth's surface showing ensemble members (grey), and estimated covariance functions (black). For computational reasons empirical covariances are shown only out to 18 degrees for the lithospheric field.  Middle row: power spectra for 5000 realizations generated from the estimated covariance functions (c) at the CMB, and (d) at Earth's surface with the CHAOS-7 model (up to degree 13) and from the LCS-1 model (above degree 16) for reference.  Bottom row: empirical probability density functions for radial magnetic field values on approximately equal area grids at (e) the CMB and  (f) Earth's surface, derived from the prior ensembles of core and lithospheric magnetic fields respectively, after transformation to an uncorrelated and normalized latent space.}
	\label{fig:hist_spect_prior_distribution}
\end{figure}

Our prior for lithospheric field comes from simulations of the lithospheric magnetization based on the forward modelling scheme developed by \citet{Hemant_Maus_2005}, with revised oceanic magnetisation according to \citet{Masterton_2013} and subduction zone magnetisations following  \citet{Williams_2019}.  We produced an ensemble of $6000$ realizations of the lithospheric field by (i) varying the crustal thickness within a range given by published crustal thickness models \citep{Nataf1996_3smac, Reguzzoni2015_gemma_crust}, (ii) varying parameters of the remanent vertically integrated magnetisation model \citep{Masterton_2013, Williams_2019}, and (iii) using stochastic perturbations generated using a Gaussian random field approach. The observed power spectra for the lithospheric field, for example from the LCS-1 model \citep{Olsen_2017}, lies within the spread of the power spectra of this ensemble \citep{Otzen_2022}. The simulated fields were generated up to spherical harmonic degree $255$, but here we only used them up to degree $120$. 

\begin{figure}
	(a) Core field prior realizations \\
 \centerline{\includegraphics[width=\textwidth]{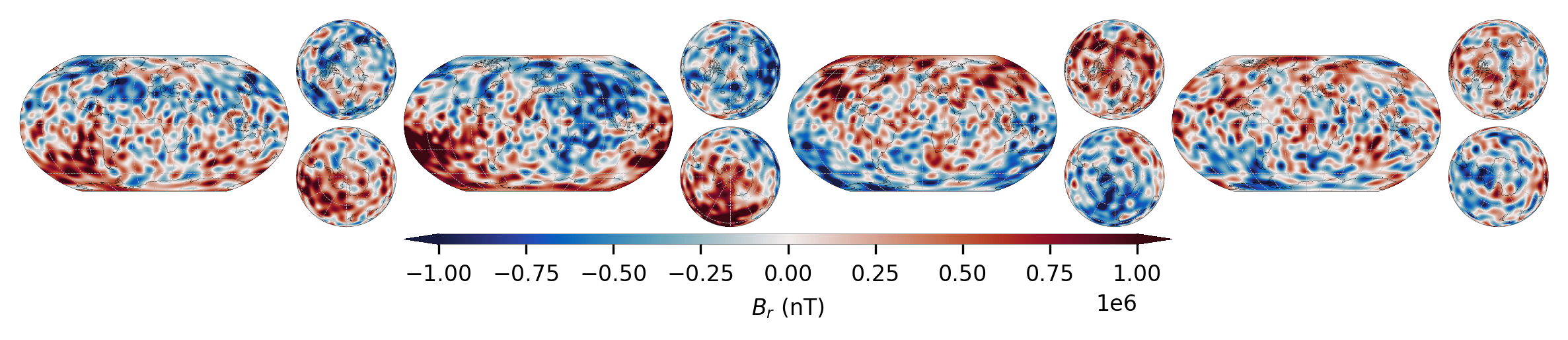}}

(b) Lithospheric field prior realizations\\
 \centerline{\includegraphics[width=\textwidth]{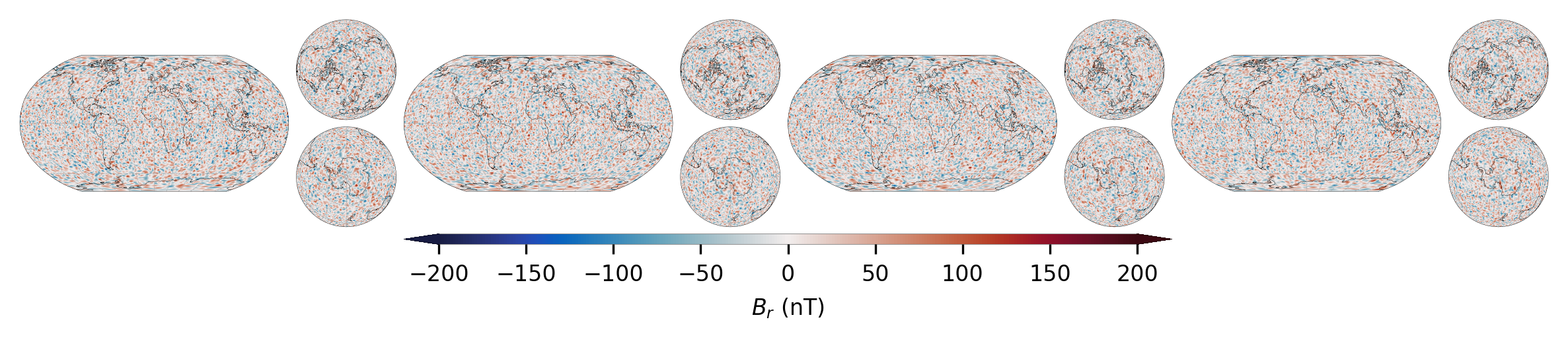}}
	\caption{Example realizations of radial magnetic fields generated using the estimated covariance functions for the core and lithospheric fields.  The core fields are shown at the core surface (top row) and the lithospheric fields are shown at Earth's surface (bottom row).  These illustrate the a-prior correlation structures assumed for the core and lithospheric fields.  Note the covariance models used to generate these prior fields are rotationally invariant.}
	\label{fig:Br_MaxEnt_prior_rand_realizations}
\end{figure}

From each ensemble member, we evaluated the radial field on an approximately equal area grid on the source surface. We used HEALPix \citep{Gorski2005_healpix} grids with $3072$ points for the core field, and $49152$ points for the lithospheric field, which are suitable for representing fields up to the spherical harmonic truncation level chosen for each source. Based on these gridded values we computed empirical semi-variograms as a function of angular distance on the spherical surface and fit covariance functions to these. We used a multi-quadratic covariance function for the lithospheric field \citep{Gneiting_2013} and a combination of a multi-quadratic covariance function and a spline function for large distances for the core field.  The resulting covariance functions, along with the empirical covariances of the ensemble members, are shown in the top row of Fig. \ref{fig:hist_spect_prior_distribution} along with their corresponding power spectra at the source surfaces generated using these covariance functions are shown in the middle row.   These covariance functions provide us with the a-priori expected covariance structure for our core and lithospheric fields. Fig. \ref{fig:Br_MaxEnt_prior_rand_realizations} shows example realizations of the core and lithospheric field generated using these covariance functions. 

We also make use of the distribution of the radial field at the sources surfaces from our core and lithospheric field ensembles, after transformation to the latent space (see equation (\ref{Eqn:latent}) and the related discussion).  These distributions and relevant statistics are presented in the bottom row of Fig. \ref{fig:hist_spect_prior_distribution}.   In particular we use expected absolute values, $<|x|>=\int |x|\, p(x) \, dx$, calculated using the mean empirical pdfs shown in Fig. \ref{fig:hist_spect_prior_distribution}, to define the latent space default parameters i.e. we set $\omega=<|x|>$, separately for the core and lithospheric fields.

The time-dependence of the core field is represented by a 6th order B-spline representation smoothed by third time derivative regularization, a standard choice in time-dependent geomagnetic field models when one wishes to study field accelerations \citep[see e.g.][]{Olsen_etal_2014}.  This is formally equivalent within the Bayesian framework to assuming a-priori that the spherical harmonic coefficients are realizations of a continuous process of the form \citep{Wahba_1990}
\begin{equation}
    \label{eq:t_dep}
    g^C_{n,m}(t) = K^{C,0}_{n,m}+ K^{C,1}_{n,m}\,t +K^{C,2}_{n,m}\,t^2 + f_{n,m}(t) ,
\end{equation}
where $f_{n,m}(t)$ is a zero mean Gaussian process \citep[see e.g.][]{Rasmussen2006} specified by the covariance function
\begin{equation}
    \label{eq:equiv_kernel}
    C_n(t_a,t_b) = \sigma_n^2 \int {(t_a-u)_+^2 \over 2}{(t_b-u)_+^2\over 2}  \, du ,
\end{equation}
where $t_a$ and $t_b$ are arbitrary times, $u$ is a dummy integration variable, and we have used the notation $(z)_+=z$ for $z \ge 0$ and $(z)_+=0$ otherwise. $K^{C,0}_{n,m}$ , $K^{C,1}_{n,m}$ and $K^{C,2}_{n,m}$ are constants associated with constant, linear and quadratic time-dependences that can be different for each spherical harmonic coefficient. $\sigma^2_n$ is the variance of the process that depends on the spherical harmonic degree $n$ and is related to the choice of regularization parameter in the standard field modelling framework. 

\subsubsection{Likelihood of geomagnetic observations}
Turning to the observations, we assume the geomagnetic measurements are contaminated by unmodelled signals that follow a long-tailed Huber error distribution.  The appropriate likelihood function is then

\begin{equation}
    P(\mathbf{d} \, \vert \, \mathbf{m}) \propto \exp\left[ -\frac{1}{2} \chi^{2}(\mathbf{m})\right]
\end{equation}
where $\chi^{2}(\mathbf{m}) = \mathbf{e}^T \, \mathbf{W} \, \mathbf{e}$ is a robust (Huber weighted) misfit norm.  $\mathbf{e}=\mathbf{d}-\mathbf{g}(\mathbf{m)}$ are the residuals between the ground and satellite magnetic observations $\mathbf{d}$ and the associated model predictions $\mathbf{g}(\mathbf{m)}$. $\mathbf{W}= \mathbf{C_e}^{-1/2}\mathbf{W_h}\mathbf{C_e}^{-1/2}$ where $\mathbf{C_e}$ is the a-priori data error covariance matrix and $\mathbf{W_h}$ a diagonal matrix that implements robust (Huber) weights and has elements $W^{i,i}_h=\mbox{min} [1, \left( c\, \sigma^d_{i}/|e_i| \right)]$ where $\sigma^d_{i}$ is the a-priori expected error for the $i$th datum and  $c=1.5$ is a constant \citep{Constable1988,Olsen_2002,sabaka2004extending}.

\subsubsection{Estimation of the model posterior probability density function}
Applying Bayes theorem the posterior probability function is
\begin{equation}
    P(\mathbf{m} \, \vert \, \mathbf{d}) \propto \exp \left[ -\frac{1}{2}\chi^{2}(\mathbf{m}) +  S_{tav} (\mathbf{x}^{C}) + S (\mathbf{x}^{L}) - \frac{1}{2} \mathbf{m}^{T} \mathbf{C}_T^{-1} \mathbf{m} \right]
\end{equation}

This can be maximized by minimizing the loss function  
\begin{equation}
 \Phi(\mathbf{m}) = \frac{1}{2}\chi^{2}(\mathbf{m}) -  S_{tav} (\mathbf{x}^{C}) - S (\mathbf{x}^{L}) + \frac{1}{2}\mathbf{m}^{T} \mathbf{C}_T^{-1} \mathbf{m}. 
\end{equation}
Here $S_{tav}(\mathbf{x}^{C})={1 \over N_P}\sum\limits_{p=1}^{N_P} S [\mathbf{x}^{C}(t_p)]$ approximates the  information entropy of the (spatially decorrelated) CMB radial field averaged over time \citep{Gillet_etal_2007} by summing the entropy at each discrete epoch $t_p$ over $N_P = 1000$ epochs.  $\mathbf{x}^{C}(t_p) = \mathbf{L}_{C}^{-1}\mathbf{G}^{C,t_p}  \mathbf{m}^{C}$ changes with $t_p$, $\mathbf{x}^{L} = \mathbf{L}_L^{-1}\mathbf{G}^L \mathbf{m}^{L}$ is assumed to be static. $\mathbf{C}_T$ is the a-prior model (temporal) covariance matrix, which includes the temporal prior information from Eqn.\,\ref{eq:equiv_kernel} for the core part of the model.   

The minimization is achieved iteratively using a Newton-type descent algorithm.  The $(k+1)$th estimate of the posterior mean model is obtained based on the model at the previous $k$th iteration by
\begin{equation}
\mathbf{m}_{k+1} = \mathbf{m}_{k} + \left[\mathbf{A}_k^T \mathbf{W}_k \mathbf{A}_k + \alphav^{C}_{k} + \alphav^{L}_{k} + \mathbf{C}_T^{-1} \right]^{-1} \cdot \left[\mathbf{A}_k^T \mathbf{W}_k \mathbf{e}_k  - \mathbf{C}_T^{-1}\mathbf{m}_{k} -
\betav^{C}_{k} - \betav^{L}_{k} \right]
\label{eq:mod_est}
\end{equation}
where $\mathbf{A}_k = (A_{ij})_k = {\partial g_i(\mathbf{m}_k) \over \partial m_j}$ is the Jacobian matrix of partial derivatives of the forward model for each datum with respect to the model parameters, evaluated using the model parameters $\mathbf{m}_k$. We have  used a notation similar to that of \citet{Stockmann_etal_2009} to define matrices, related to the Hessian matrix of the second-order partial derivatives of the entropy function, of the form
\begin{equation}
\alphav=\left(\mathbf{L}^{-1}\mathbf{G}\right)^{T} \mathbf{S}_H \left(\mathbf{L}^{-1}\mathbf{G}\right) \quad  \mbox{where} \quad  \mathbf{S}_H = \, \mbox{diag}\left[ {4 \omega \over \psi_1},{4 \omega \over \psi_2},..., {4 \omega \over \psi_M}  \right]^T,
\label{eq:alphav}
\end{equation}
and vectors, related to the first order partial derivatives of the entropy function, of the form
\begin{equation}
\betav=  \left(\mathbf{L}^{-1}\mathbf{G}\right)^{T} \mathbf{S}_G  \quad  \mbox{where} \quad  \mathbf{S}_G = 4\omega \left[ \log \left({\psi_1 + x_1 \over 2 \omega}\right),\log \left({\psi_2 + x_2 \over 2 \omega}\right),..., \log \left({\psi_M + x_M \over 2 \omega}\right)  \right]^T.
\label{eq:betavC}
\end{equation}

In $\alphav^C_k$, $\betav^C_k$ and $\alphav^L_k$ , $\betav^L_k$ in Eq.\,(\ref{eq:mod_est}) the superscripts $C$ and $L$ refer to the core and lithospheric fields respectively, while subscript $k$ denotes that the computation is carried out using model parameters from the previous $k$\,th iteration. 
In $\alphav^{C}$ and $\betav^{C}$ the expressions for $\alphav$ and $\betav$ at each epoch $t_p$ must be averaged over time in the same way as $S_{tav}$ is defined above.

Minimizing measures of the data misfit and the temporal complexity while maximizing the entropy of the (decorrelated) core and lithospheric radial fields at their source surfaces results in internal fields that fit the observations and are compatible with the temporal prior but allow high dynamic ranges of $\mathbf{x}$ at the source surfaces while satisfying the spatial covariance properties of the core and lithospheric priors.

After convergence of the above scheme we describe the dispersion of the posterior distribution using an approximate model covariance matrix defined about the maximum of the posterior pdf, computed from the Hessian of the loss function $\Phi(\mathbf{m})$ by \citep[e.g.][]{Tarantola_2005,Hobson_etal_1998} 
\begin{equation}
\mathbf{C_m} \approx (\mathbf{A}^T \mathbf{W} \mathbf{A} + \mathbf{\alphav^{C}} + \mathbf{\alphav^{L}} + \mathbf{C}_T^{-1})^{-1}
\label{eq:mod_cov}
\end{equation}
where $\mathbf{A}$, $\mathbf{W}$, $\mathbf{\alphav^{C}}$ and $\mathbf{\alphav^{L}}$ are the values of $\mathbf{A}_k$ $\mathbf{W}_k$, $\mathbf{\alphav^{C}_k}$ and $\mathbf{\alphav^{L}_k}$ from the final iteration, when the scheme is considered to have converged.

%******************************************

\section{Observations}
\label{sec:data}
The models reported here are built from a dataset of geomagnetic observations similar to that used to construct the CHAOS-7 geomagnetic field model \citep{Finlay_CHAOS7:2020}, but restricted to the period between 2005 and 2020.

\begin{figure}
    \centerline{\includegraphics[width=\textwidth]{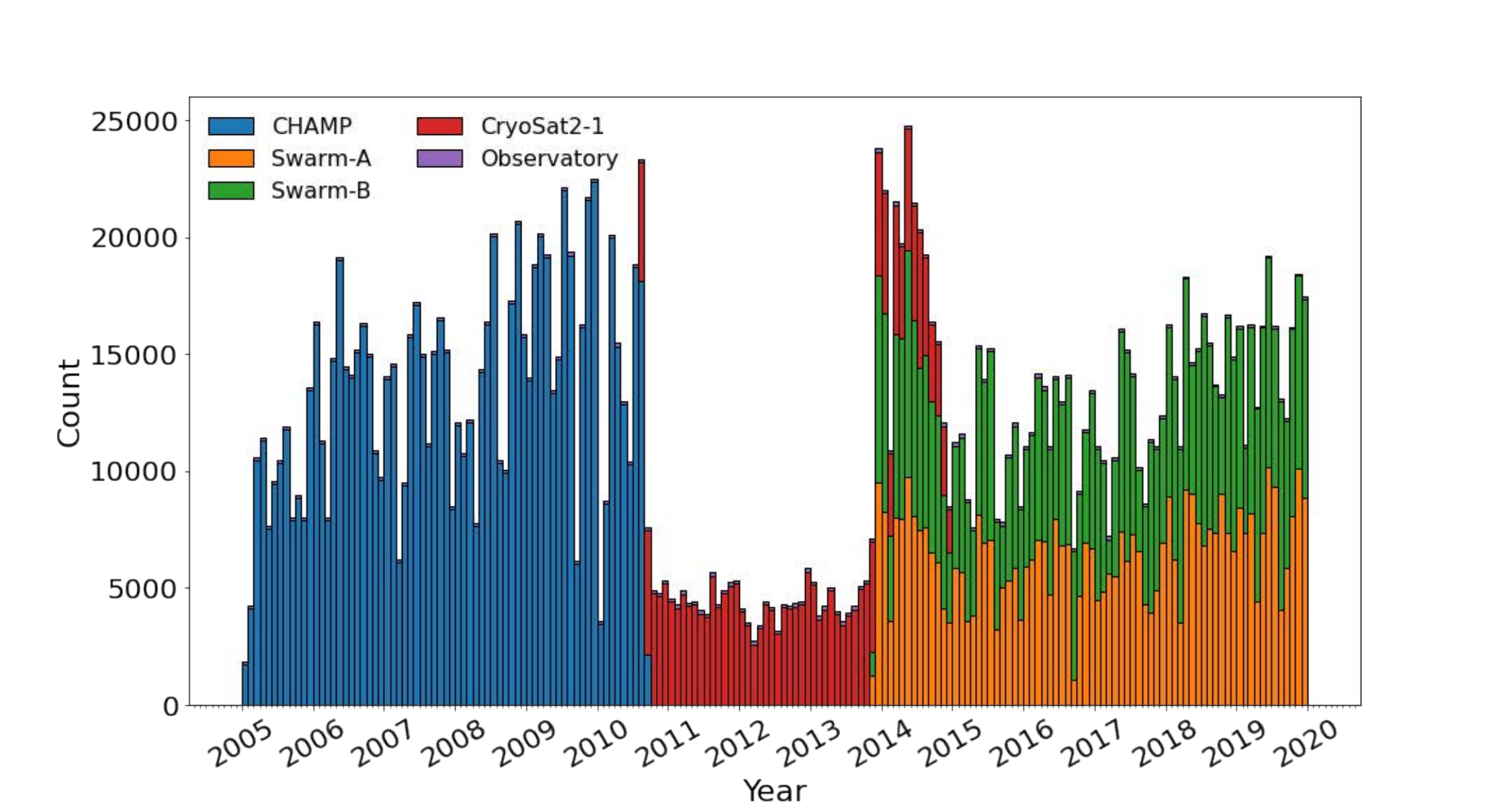}}
	\caption{Stacked histogram showing number of vector field observations employed per month between 2005 and 2020.  Colours indicate the data source.}
	\label{fig:data_distribution}
\end{figure}

Observations from the CHAMP, Cryosat-2 and \textit{Swarm} A and B satellites are used, three-component vector measurement for quasi-dipole latitudes up to 55 degrees and scalar intensity data at higher latitudes.  Level 3 CHAMP magnetic field data, Cryosat-2 L1b magnetic field data (FGM 1, the dataset used in the CHAOS-7 model, here pre-calibrated using CHAOS-7) and \textit{Swarm} L1b magnetic field data (version 0601) are used with 1 minute sampling for CHAMP and Cryosat-2 and 2 minute sampling for each \textit{Swarm} satellite.  We also used along-track gradients from CHAMP, \textit{Swarm} A and \textit{Swarm} B; gradients are particularly useful for constraining the high degree lithospheric field.  Geomagnetic quiet-time selection criteria were employed such that Kp $ \le 2^0$, d$|$RC$|$/dt  $\le$ 2\,nT/hr \citep{Olsen_etal_2014}, averaging over the previous 2 hours the merging electric field at the magnetopause $E_m \le 0.8$\,mV/m, the interplanetary magnetic field (IMF) $B_Z>0$ , and IMF $B_Y$ is less than\,3 nT in the northern hemisphere and greater than -3\,nT in the southern hemisphere. Only data from dark conditions (sun at least 10 degrees below the horizon) were used.  A more detailed description of these data selection criteria is found in \citet{Finlay_CHAOS7:2020}.  In addition to satellite observations, as in CHAOS-7 we used annual differences of revised monthly means from ground observatories are used, as in the CHAOS-7 model, based on hourly mean values from the BGS AUX$\_$OBS database, version  0129.  A stacked histogram of the number of vector field observations used versus time is presented in Fig.\,\ref{fig:data_distribution}.

%******************************************
\section{Results}
\label{sec:results}

\subsection{Implementation and model diagnostics}
\label{sec:core}

We now briefly document here some details regarding the practical implementation before moving on to the results.  In order to make our modelling setup as close as possible to the CHAOS field modelling scheme, in $ \mathbf{C}_T^{-1} $  we consider sub-matrices associated with calculating quadratic norms of (i) the 3rd time derivative of the internal radial field integrated over the core surface and and throughout the model time span (formally equivalent to the the prior defined in Eqn.\,\ref{eq:t_dep}) (ii) the acceleration of the core field at the model endpoints and (iii) temporal first differences of the estimated offsets of the external dipole in solar-magnetic coordinates (related to imperfections in the RC index). The related  hyperparameters denoted by $\lambda_{i3}$, $\lambda_{i2e}$, and $\lambda_{sm}$ are fixed throughout this study and implicitly included within $\mathbf{C}_T^{-1}$.  They were chosen so as to produce time-variations similar to the CHAOS-7 field model although for simplicity we did not use degree-dependent tapering or treat zonal terms differently.  The relevant hyperparameters related to the entropy default parameters for the dimensionless latent variables, $\omega^C$ and $\omega^L$, were set to values 0.412 and 0.422 based on the expected absolute values of $\mathbf{x}^{C}$ and $\mathbf{x}^{L}$ from the distributions of the dimensionless latent variables found in the prior ensembles (see Fig.\,\ref{fig:hist_spect_prior_distribution}, bottom). The adopted hyperparameters are collected in table \ref{table.1}.

\begin{table}
\caption{\textbf{Modelling hyperparameters}}
\centering
{\small
\begin{tabular}{p{9cm} p{5cm}}
\toprule
  \textbf{Parameter} & \textbf{Value} \\
\midrule
 $\omega^C$ & $0.412$\\
 $\omega^L$ & $0.422$\\
 $\lambda_{i3}$ & $1$ $(\text{nT yr}^{-3})^{-2}$ \\
 $\lambda_{i2e}$ & $100$ $(\text{nT yr}^{-2})^{-2}$ \\
 $\lambda_{sm}$ & $1200$ $(\text{nT yr}^{-1})^{-2}$\\
\bottomrule                
\end{tabular}
\label{table.1}}
\end{table}

 The full estimated model, including the time-dependent core field to degree 30, static lithospheric field to degree 120, external field parameters and alignment parameters for each satellite, consists of $49495$ parameters in all. We started the iterative model estimation scheme with model parameters for the core taken from CHAOS-6.9 up to degree $n=12$, and for the lithosphere from the LCS-1 model at degree $n=16$ and above. The small-scale core field and large-scale lithospheric field were otherwise initialized with zeros.  After  $24$ iterations the largest change in a model parameter relative to its amplitude was $0.0037\%$ and no further change was seen in CMB maps of the posterior mean core field. Below we refer to the resulting model, including co-estimated core and lithospheric parts, as model CL. 

 For comparison, we also built a more traditional CHAOS-type field model, with a single time-dependent internal field up to spherical harmonic degree 20 and a static internal field for degrees 21 to 120, using the same external field parameterization and hyperparameters, covering the same period, and based on the same dataset as used to build the model CL.  For this model we considered the estimated core field to be the time-dependent internal field up to degree 13, as has been the standard practise when interpreting the CHAOS model \citep[e.g.][]{Olsen_etal_2014}.

Table \ref{table.2} collects Huber-weighted means and RMS residuals between the vector field data and the model predictions (in nT), comparing model CL and our CHAOS-type reference model. Model CL fits the satellite and ground data overall to a similar level as the CHAOS-type model, while simultaneously minimizing the information entropy of the (spatially-decorrelated) time-dependent core and lithospheric fields.  
The two models are found to have very similar temporal regularization norms, which is not surprising at they were built using the same temporal hyperparameters.  The non-dimensional information entropy norms for the decorrelated core and lithospheric fields, $S^{C}_{tav}$ and $S^{L}$, after 24 iterations were respectively $0.94$ and $5.31$ for the CL model.

\begin{table}
\caption{\textbf{Misfit statistics for vector field data in the non-polar region and scalar data in the polar region. Note that Huber weights were employed when calculating the reported statistics. (Misfits for gradient data not shown).}}
\centering
{\small
\begin{tabular}{p{2.9cm} p{1.7cm} p{2cm} p{1.4cm} p{1.4cm} p{1.4cm} p{1.4cm}}
\toprule
& & & \multicolumn{2}{c}{\textbf{CL}} & \multicolumn{2}{c}{\textbf{CHAOS-type}}\\
\textbf{Source} & \textbf{QD lat} & \textbf{N} & \textbf{Mean (nT)} & \textbf{RMS (nT)} & \textbf{Mean (nT)} & \textbf{RMS (nT)}\\
\midrule
CHAMP     & non-polar & $2,080,146$ & $-0.003$ & $2.397$  & $-0.003$  & $2.397$\\
          & polar     & $98,751$    & $-0.001$ & $3.498$  & $-0.004$  & $3.496$\\
CryoSat-2 & non-polar & $1,017,960$ & $-0.005$ & $6.160$  & $-0.012$  & $6.160$\\
          & polar     & $59,509$    & $1.371$  & $6.556$  & $1.538$   & $6.604$\\
Swarm-A   & non-polar & $1,085,118$ & $0.007$  & $2.182$  & $0.007$   & $2.182$\\
          & polar     & $47,936$    & $0.039$  & $3.117$  & $0.045$   & $3.117$\\
Swarm-B   & non-polar & $1,079,508$ & $-0.016$ & $2.176$  & $-0.016$  & $2.176$\\
          & polar     & $48,433$    & $0.194$  & $2.905$  & $0.203$   & $2.906$\\
\midrule
& & & \textbf{(nT/yr)} & \textbf{(nT/yr)} & \textbf{(nT/yr)} & \textbf{(nT/yr)}\\
\midrule
Ground observatory & & $67,632$ & $-0.027$ & $3.304$ & $-0.026$ & $3.306$\\
\bottomrule                
\end{tabular}
\label{table.2}}
\end{table}

\newpage

\subsection{The core-mantle boundary field}
\label{sec:core}

\begin{figure}
	\centerline{\includegraphics[width=0.9\textwidth]{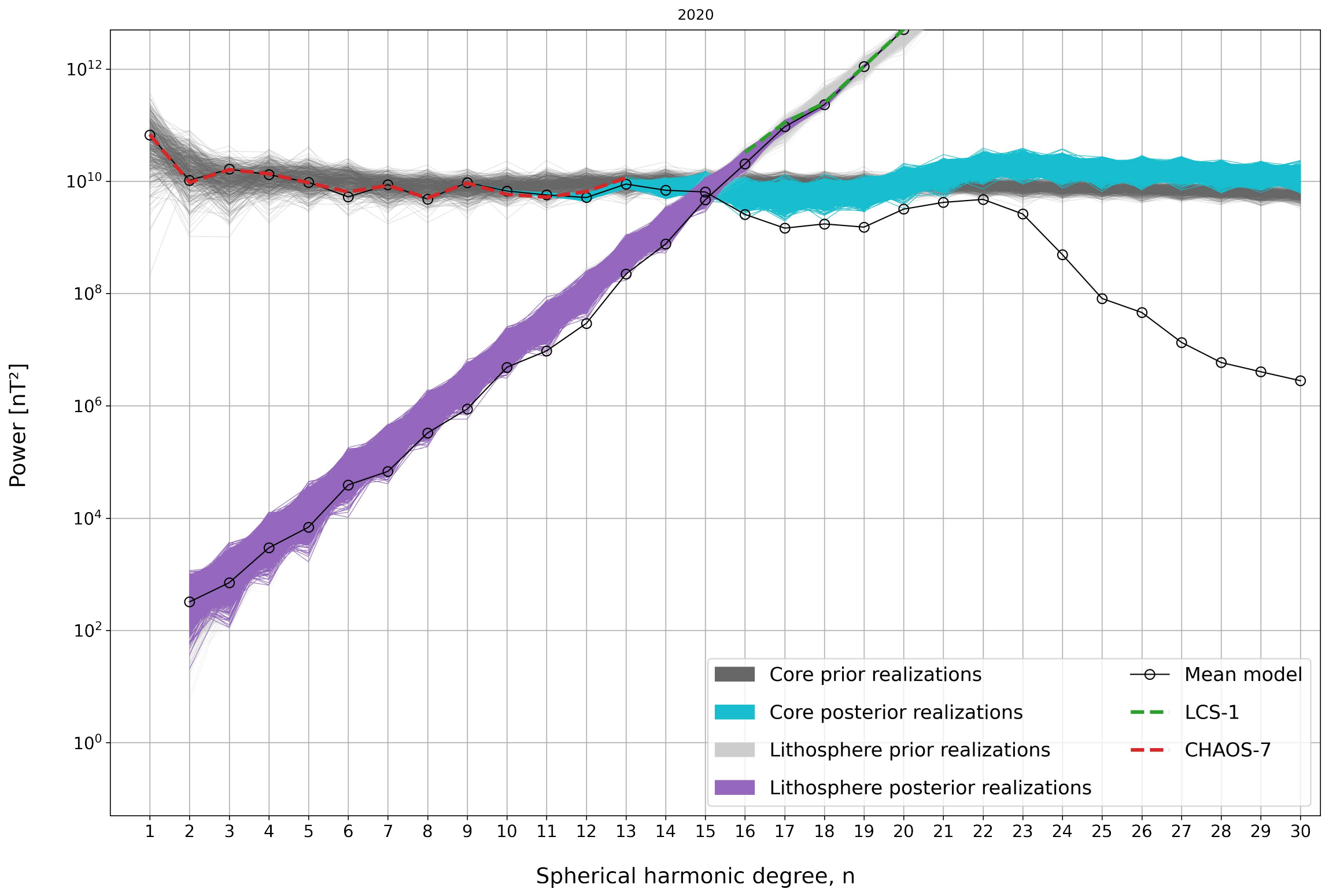}}
	\caption{Lowes-Mauersberger spectra showing power as a function of spherical harmonic degree at the core-mantle boundary in 2020.0 for model CL. Solid black lines with circles show the estimated posterior mean core and lithospheric field models which cross at degree 15. Grey lines show realizations based on the adopted prior covariance model, cyan lines show posterior realizations of the core field and purple lines posterior realizations of the lithospheric field. Degree 1 is not shown for the lithospheric field as it is not well separated from the core field. CHAOS-7 (up to degree 13) and LCS-1 (at degree 16 and above) are shown for reference. }
	\label{fig:pspec_spatial_cmb}
\end{figure}

The spatial power spectra \citep{Lowes:1966,Mauersberger:1956} of model CL at the CMB in 2020 is presented in Fig.\,\ref{fig:pspec_spatial_cmb}. The obtained posterior mean model closely follows the internal field from CHAOS-7 up to degree 11 but contains slightly less power at degree 12 and 13. Here and below comparisons to CHAOS-7 used version CHAOS-7.9.  Internal field models (such as the CHAOS model) are usually truncated at degree 13 when carrying out interpretations at the CMB since above this degree their CMB spectra diverge as they also include signals from the lithospheric field.   The posterior realizations from the model CL core field have power spectra that are approximately flat out to degree 30 and do not diverge. The posterior mean model shows a gradual drop in power for degrees 15 to 19 and a slight increase again for degrees 20 to 22.  Above degree 22 the power in the posterior mean model drops to much lower levels indicating that field realizations essentially average to zero at higher degrees where almost all the observed signal comes from the lithospheric field.  

The power in the lithospheric field realizations and mean models mapped down to the CMB is also shown in Fig.\,\ref{fig:pspec_spatial_cmb}, these do diverge.  The core and lithospheric field spectra cross between degree 14 and 16 for realizations of model CL, and at degree 15 for the mean models.  Note that the estimated lithospheric field is presented only for degree 2 and above, at degree 1 it is not well separated from the core field, which we believe is a consequence of the entropy function of the decorrelated latent variables not being greatly affected by changes in the dipole field. 

\begin{figure}
	\centerline{\includegraphics[width=\textwidth]{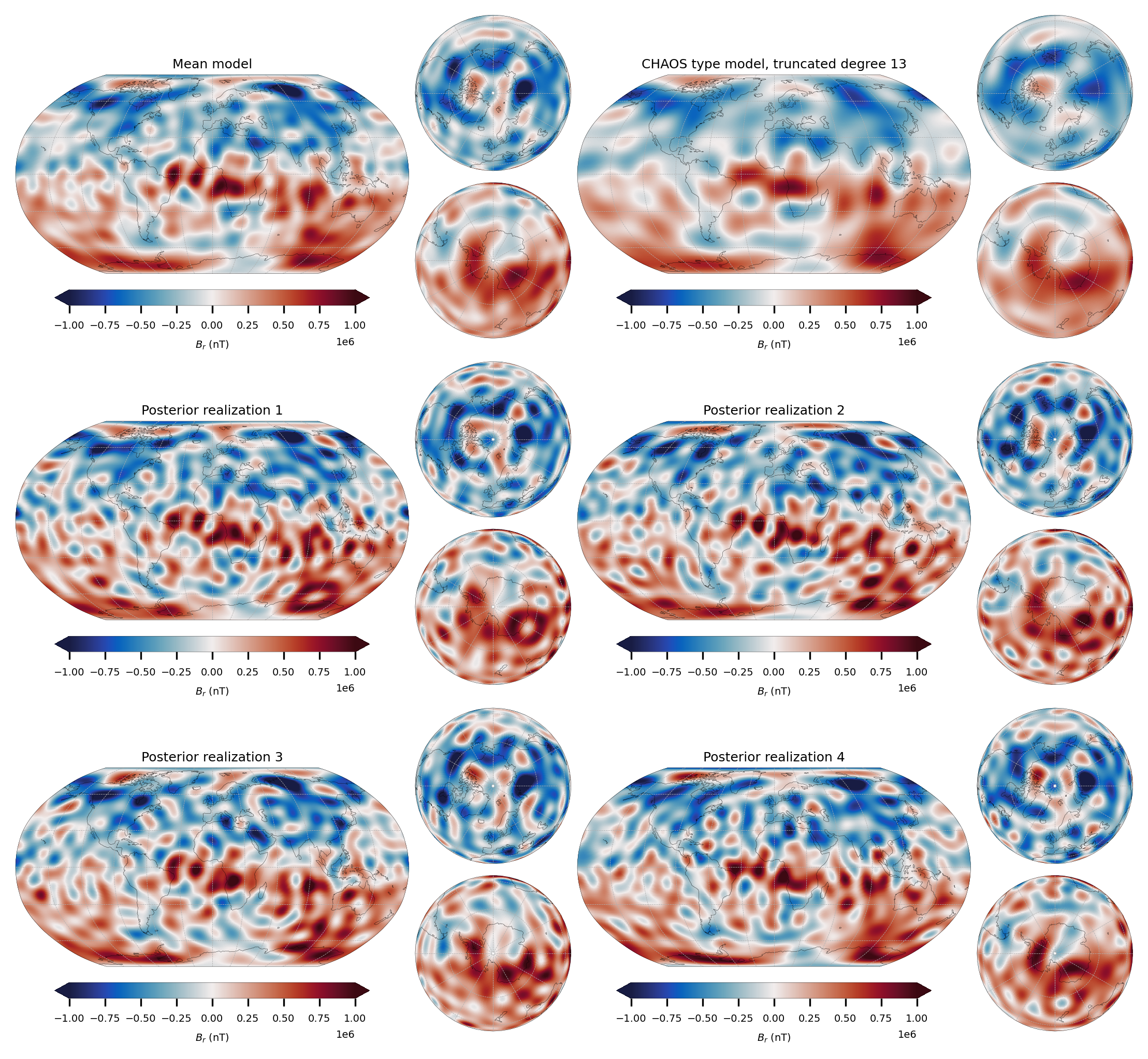}}
	\caption{Radial magnetic field maps at the core-mantle boundary in 2020 for model CL. Estimated posterior mean model (top left), a CHAOS-type model constructed from the same dataset (top right), and four example posterior realizations.  All models are truncated at degree 22 except for the CHAOS-type model which is truncated at degree 13. The common features across all posterior models, as captured in the mean model, provide information on the core field to beyond degree 13.}
	\label{fig:Br_MaxEnt_core_psr}
\end{figure}

Maps of the radial magnetic field at the CMB in 2020.0 from model CL, for the posterior mean core field model and four example posterior realizations all truncated at degree 22, are presented in Fig.\,\ref{fig:Br_MaxEnt_core_psr}. A similar map from the CHAOS-type reference model, truncated in the conventional fashion at degree 13, is shown for reference. The posterior realizations contain more power at small length scales, but all realizations agree on the larger-scale structure as represented by the mean model.  

\begin{figure}
	\centerline{\includegraphics[width=1.0\textwidth]{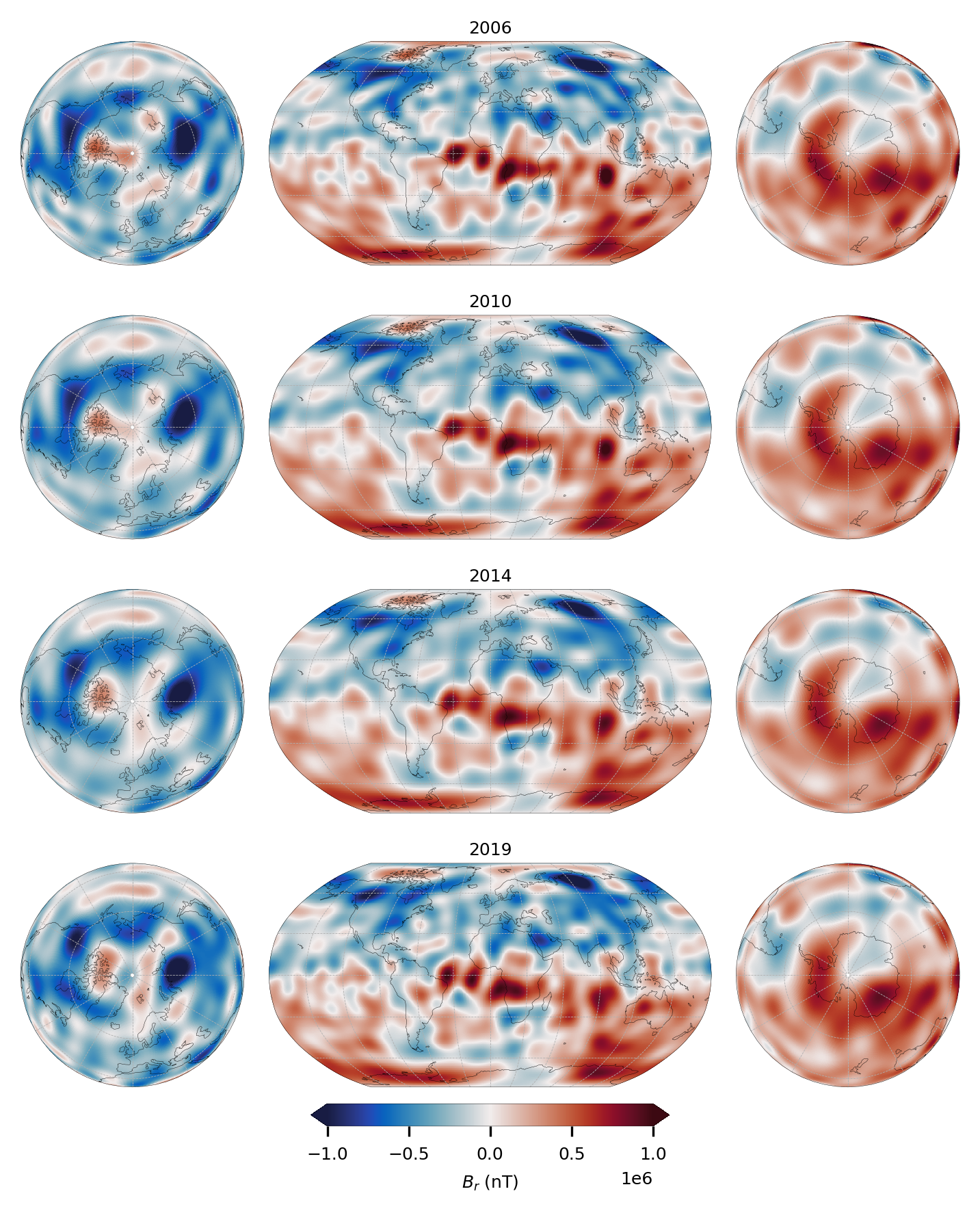}} 
	\caption{Time sequence of maps of radial field at the core-mantle boundary in 2006, 2010, 2014 and 2019 from the model CL posterior mean truncated at degree 22.  Changes are due both to the true evolution of the core field and changes in observational constraints across the epochs.}
	\label{fig:Br_MaxEnt_core_t_evo}
\end{figure}

More details of the CMB field structures are evident in the model CL posterior mean compared to the traditional CHAOS-type model truncated at degree 13. Some low latitude flux concentrations are split, see for example the two strong positive radial field features near the equator between Africa and South America which are usually interpreted as single feature (as in the CHAOS-type model). A strong high latitude flux feature under Siberia, located under the Taymyr peninsula in central northern Siberia in 2020 is found to be more localized and stronger than in models truncated at degree 13.  This feature has moved north-westwards between 2005 and 2020, as seen in Fig. \ref{fig:Br_MaxEnt_core_t_evo} which shows the posterior mean map up to spherical harmonic degree 22 at a sequence of times.   

Concerning reversed flux patches in the South Atlantic, we find evidence for two reversed patches under South Africa adjacent to strong norm flux patches under central Africa (see also their time evolution in Fig.\, \ref{fig:Br_MaxEnt_core_t_evo}).  Regarding the reverse flux region under the Southern Atlantic ocean, there are several distinct reversed flux concentrations visible within this region, which are observed to evolve separately.  

\begin{figure}
    (a) \qquad \qquad \qquad \qquad \qquad \qquad \qquad \qquad \qquad \qquad \qquad  (b)\\
    \includegraphics[width=0.49\textwidth]{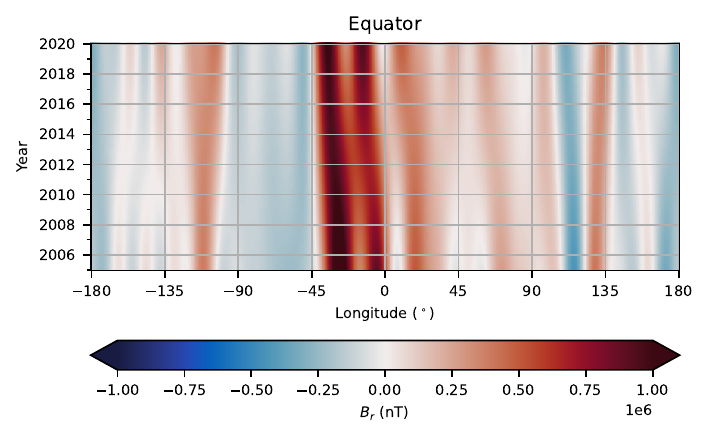} \includegraphics[width=0.49\textwidth]{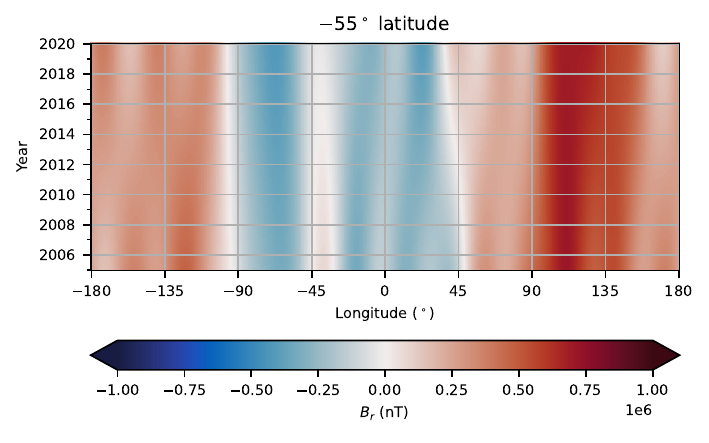}
    \caption{Time-longitude plots of the posterior mean from model CL truncated at degree 22, showing the evolution of the radial magnetic field at core-mantle boundary (a) on the equator and (b) at latitude 55 degrees south (right).}
    \label{Fig:TL}
\end{figure}

It is also evident in Fig\,\ref{fig:Br_MaxEnt_core_t_evo} that there is more power in the mean model at degrees 16 to 22 for the first five years and last six years of the model, when CHAMP and \textit{Swarm} observations respectively were available.  Fig. \ref{Fig:TL} further illustrates the time-dependence of the core field, focusing on coherent east-west motions of flux features in time-longitude plots of the CMB radial field at the equator and at 55 degrees south.  Intense equatorial features are observed to drift coherently westwards under the mid-Atlantic over the 15 years studied. There is also evidence for eastward drift of a reversed flux feature under the Southern mid-Atlantic ocean at 55 degrees south heading from South America toward Africa.  Such coherently drifting flux features involving power above spherical harmonic degree 13 cannot be ascribed to lithospheric sources. 

\subsection{The lithospheric field at Earth's surface}
\label{sec:lith}

Although not our main focus here, we present for reference details of our co-estimated lithospheric field, which includes an estimate of the large scale of the lithospheric field which is usually neglected.   

\begin{figure}
 \centerline{\includegraphics[width=0.9\textwidth]{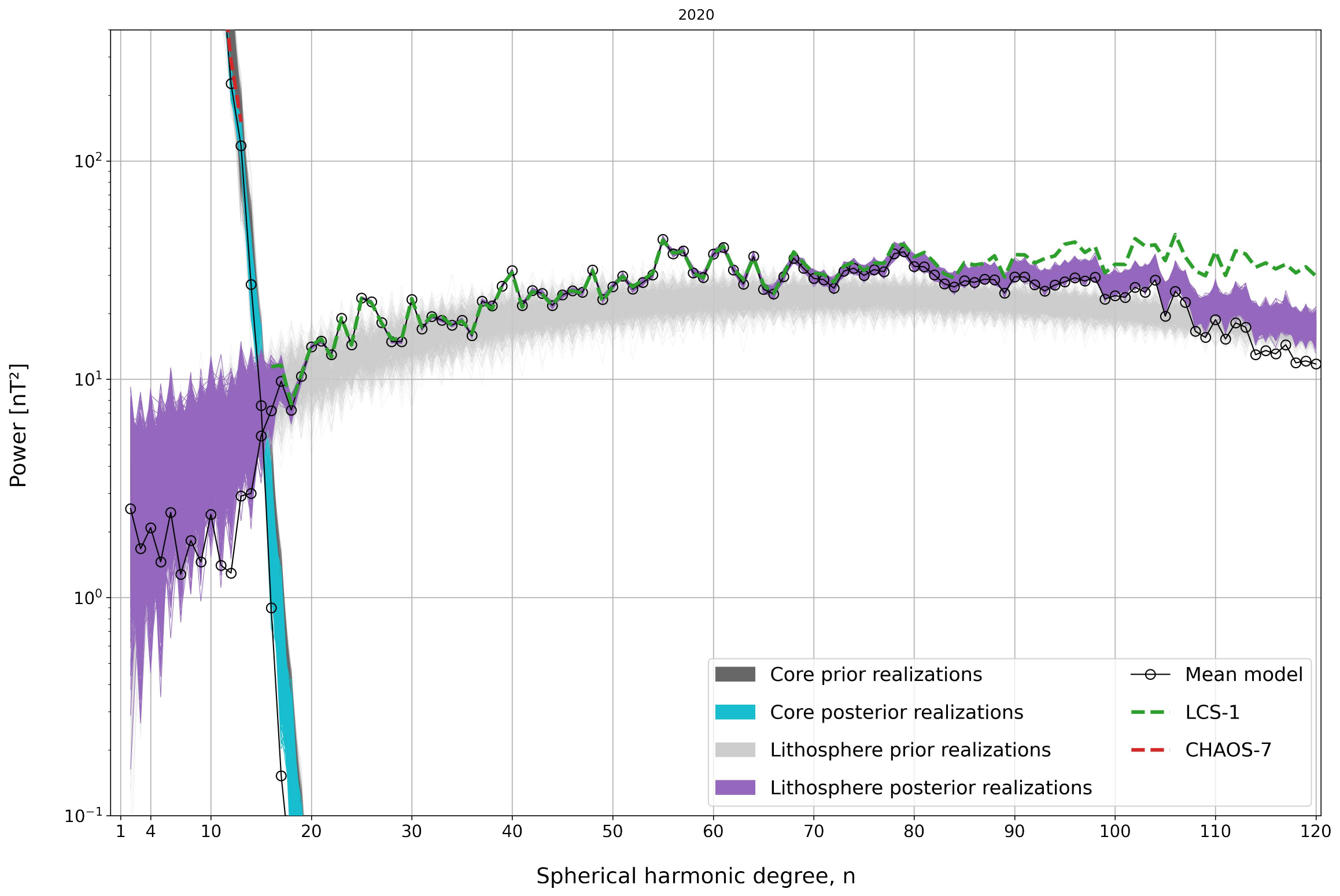}}
 \caption{Lowes-Mauersberger spectra showing power as a function of spherical harmonic degree at Earth's mean spherical reference radius in 2020.0 for model CL. Solid black lines with circles show the estimated posterior mean core and lithospheric field models which cross at degree 15. Grey lines show prior realizations based on the adopted spatial covariance model, cyan lines show posterior realizations of the core field and purple lines posterior realizations of the lithospheric field.  CHAOS-7 (up to degree 13) and LCS-1 (at degree 16 and above) are shown for reference.}
	\label{fig:Br_MaxEnt_lith_spect}
\end{figure}

Fig.\ref{fig:Br_MaxEnt_lith_spect} shows the spatial power spectra of the co-estimated core and lithospheric fields from model CL at the Earth's  mean spherical reference surface.  The spread in the posterior realizations is very small (almost invisible in the plot) at degrees 17 up to 70, indicating that the lithospheric field is very well constrained by the observations for these degrees.  At lower degrees the posterior spread increases, with the mean model containing lower power than most of the individual realizations.  The posterior spread also increases above degree 70, becoming as large as the spread in the prior above degree 110 by which point the mean model contains less power than any of the posterior realizations.  The cross-over between the mean core and lithospheric field models also occurs at degree 15 at Earth's surface.

\begin{figure}
 \centerline{\includegraphics[width=\textwidth]{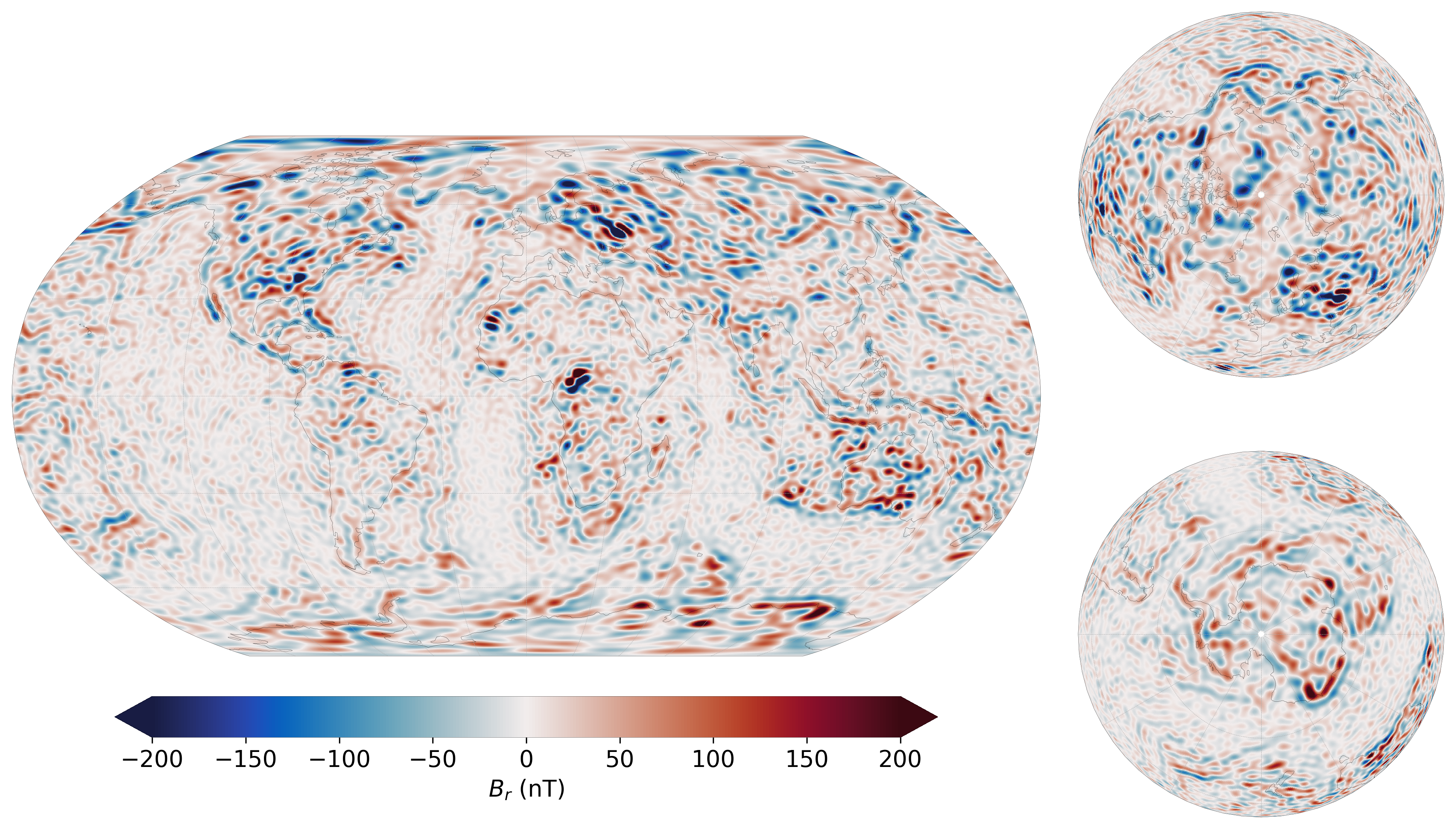}}
	\caption{Map of estimated posterior mean lithospheric field at Earth's spherical reference radius, for degrees 2 to 120.}
	\label{fig:Br_MaxEnt_lith_map}
\end{figure}

In Fig. \ref{fig:Br_MaxEnt_lith_map} we present a map of the posterior mean lithospheric field from model CL at Earth's surface for degree 2 to 120.  

\newpage

\subsection{Modelled secular variation and comparisons with ground observatories}
\label{sec:time_dep}

\begin{figure}
	\centerline{\includegraphics[width=\textwidth]{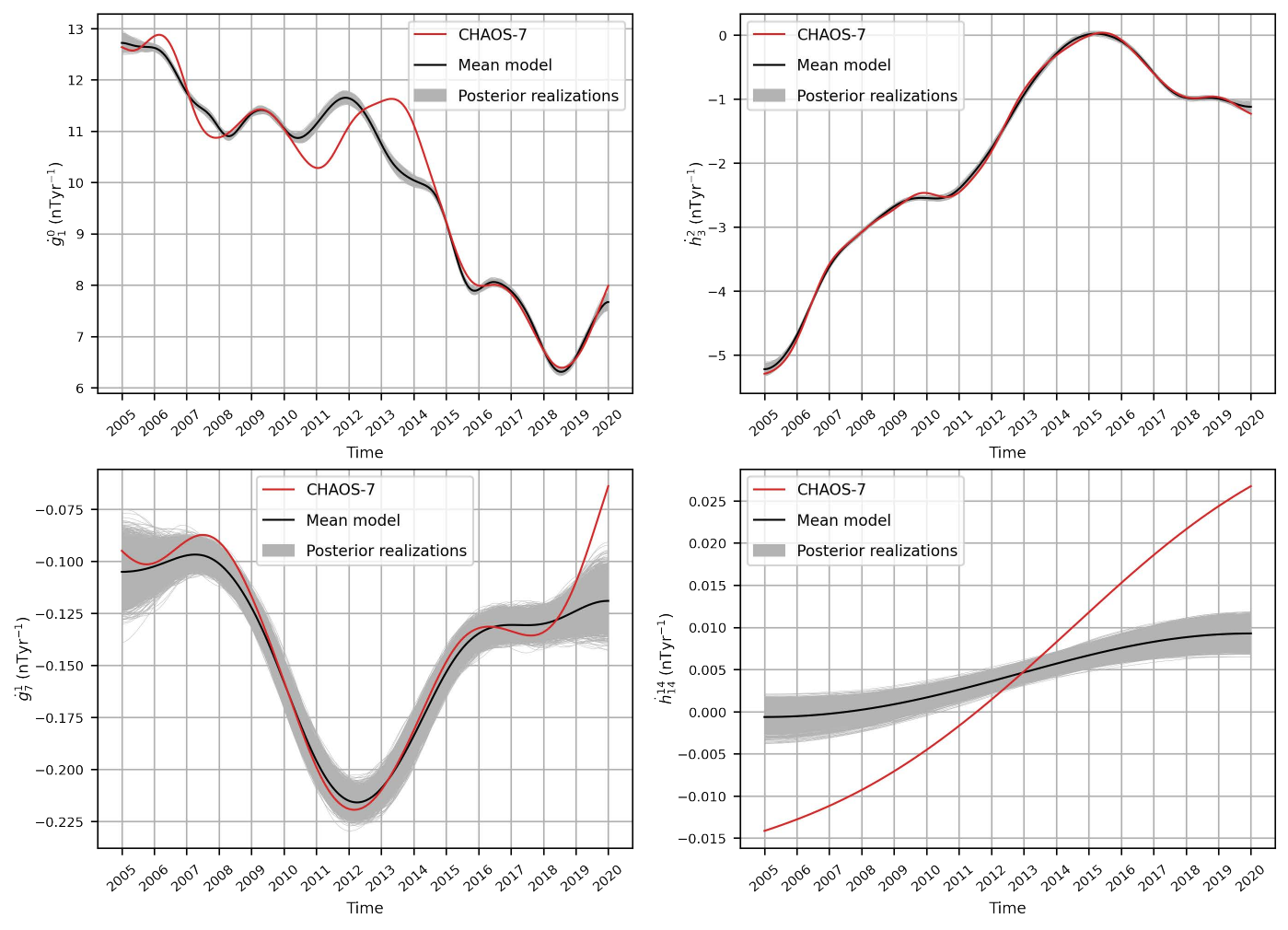}}
	\caption{First time derivative (secular variation) of selected spherical harmonic coefficients of the core field over the model timespan from 2005 to 2020. Top left: $dg^0_1/dt$, top right: $dh^2_3/dt$, bottom left: $dg^1_7/dt$ and bottom right: $dh^{14}_{14}/dt$.  The estimated posterior mean model is shown as the black solid line, posterior realizations in grey and for reference CHAOS-7 in red. }
	\label{fig:core_timeseries_gauss_SV}
\end{figure}

To document the time dependence in model CL and show that this is also reasonable, Fig.\,\ref{fig:core_timeseries_gauss_SV} presents example time-series of the first time derivative (or secular variation, SV) of spherical harmonic coefficients from model CL's core field model. It shows the posterior mean, 5000 posterior realizations, and the CHAOS-7 model for reference.  The time-dependent SV of $g^0_1$ is slightly smoother in model CL than in CHAOS-7, agrees well at intermediate degrees, following the same trends and showing some (expected) differences close to the endpoints.  Model CL generally shows lower amplitude changes in SV at high degree.  This behaviour is expected because the CHAOS-7 model tapered its regularization to lower values at high degrees.  The dispersion of the posterior realizations also increases with spherical harmonic degree.  We see no evidence for unrealistic features in the SV coefficients of model CL.

\begin{figure}
	\centerline{\includegraphics[width=\textwidth]{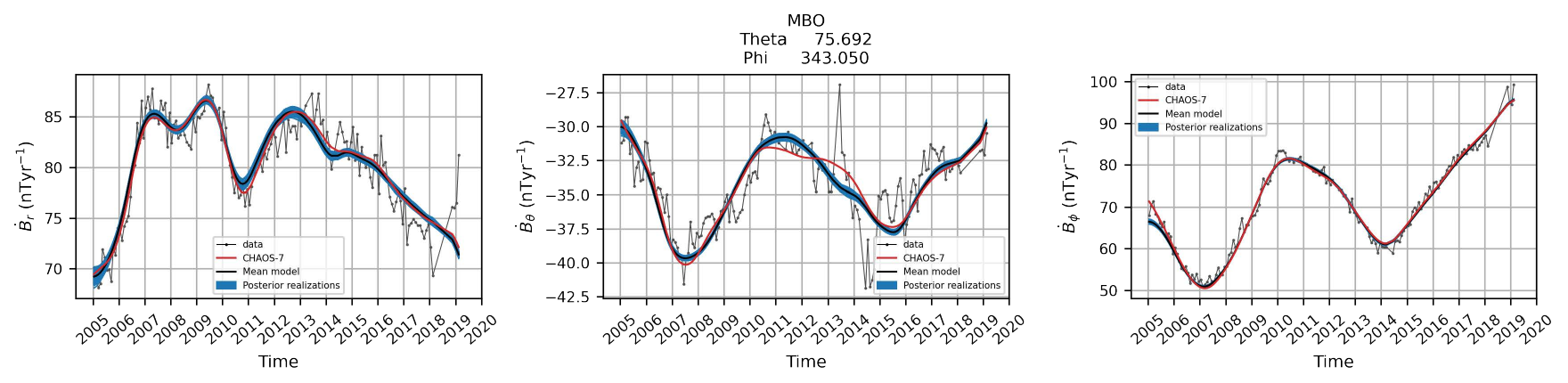}}
 
    \centerline{\includegraphics[width=\textwidth]{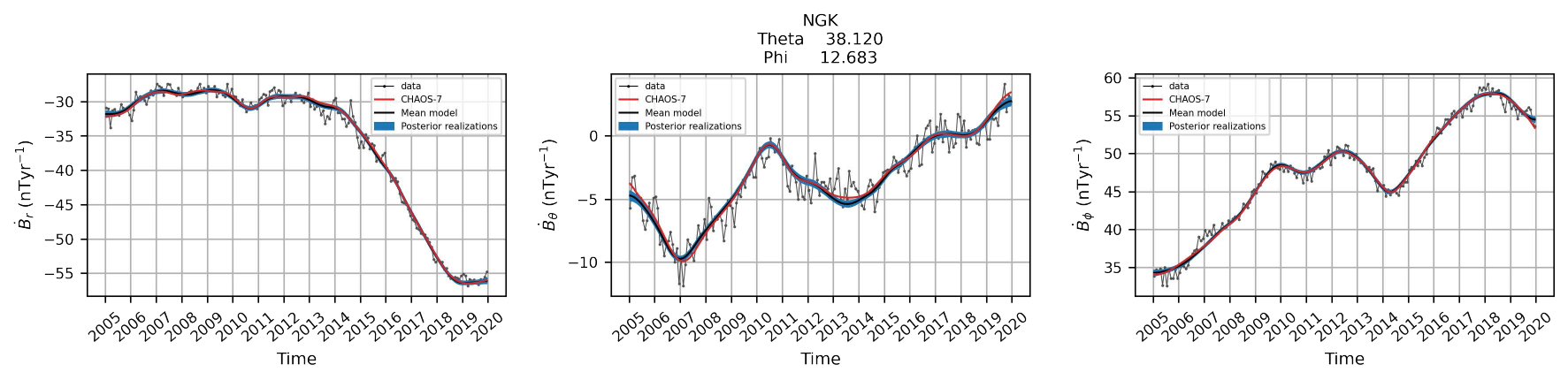}}
 
    \centerline{\includegraphics[width=\textwidth]{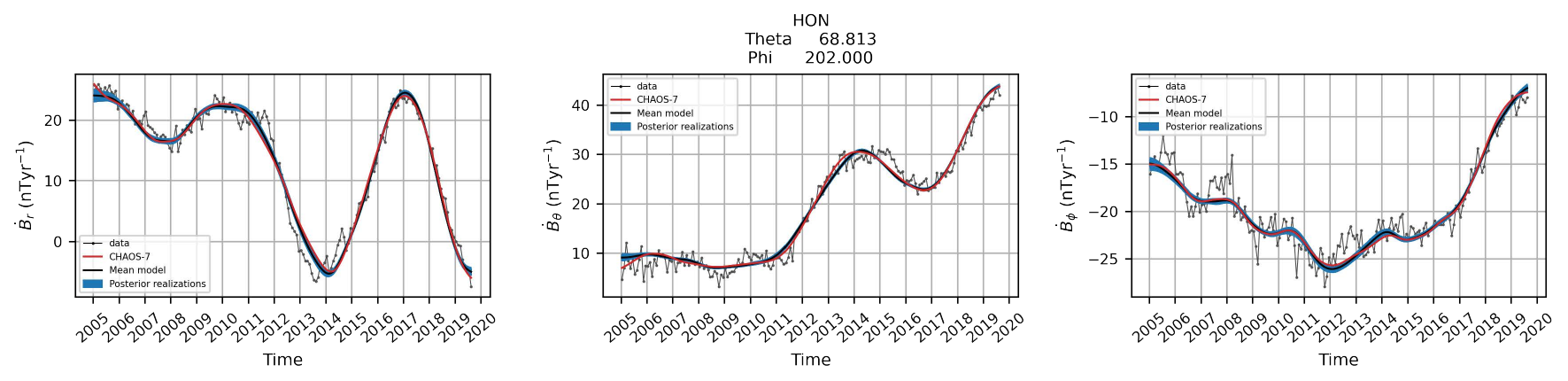}}
	\caption{Comparison of secular variation predicted by model CL for the posterior mean model (black solid line) and posterior realizations (blue lines) with annual differences of revised monthly means at selected ground observatories (thin black line with dots). CHAOS-7 is shown for reference in red.  Left column shows the time derivative of the radial field, middle column the time derivative of the southward field component and right column the time derivative of the eastward field component. Top row: M'bour observatory (MBO) from low latitudes in west Africa, middle row: Niemegk observatory (NGK) from mid-latitudes in Europe and bottom row: Honolulu observatory (HON) from mid-latitudes in the Pacific.}
	\label{fig:core_timeseries_obs}
\end{figure}

This conclusion is supported by comparisons of SV from model CL with observed SV data at ground observatories, for example as shown in Fig.\,\ref{fig:core_timeseries_obs}.  Model CL predictions agree well with the annual differences of monthly means for stations from Africa (M'Bour), Europe (Niemegk) and in the Pacific (Honolulu), with the fit being similar to that of CHAOS-7 although with somewhat smoother time variations. The posterior realizations give an indication of the formal uncertainties in the SV predictions of model CL.

\section{Discussion and Conclusions}
\label{sec:Disc_conc}

\subsection{Insights from a synthetic test}
A key question is the extent to which our method is able to retrieve the core field above degree 13. To investigate this issue we carried out a synthetic test where the true core and lithospheric fields were known. We took as input a time-dependent core field up to spherical harmonic degree 30 from a dynamo simulation, along with a synthetic lithospheric field based on a simulation of the induced and remanent lithospheric magnetisation up to degree 120.  Magnetic data were synthesized at the same locations, and for the same field components, as in the observed dataset and this was inverted using the maximum entropy co-estimation scheme described in \ref{sec:method}.  Full details and the results from this synthetic test are collected in Appendix \ref{sec:AppA}.  The synthetic data used in the test is consistent with our prior, however the utilized prior information is rather weak (involving only the source radii and isotropic spatial covariance functions); the main purpose of the test is to investigate what level of separation of the core and lithospheric fields is possible using our approach. 

Fig.\,\ref{fig:pspec_spatial_synth} presents the resulting spatial spectra at the CMB and at Earth's surface while Fig.\,\ref{fig:Br_MaxEnt_range_compare_true}  compares side-by-side the radial field at the CMB from the estimated posterior mean core field model and the input dynamo field for increasing truncation degrees of 13, 16, 19 and 22.  The posterior mean model obviously has less power than the dynamo synthetic truth from degree 16 to 22 with the missing power at small scales particularly obvious at low latitudes where the dynamo solution is most complex.  

Despite under-estimation of the power at degrees 16 to 22, Fig.\,\ref{fig:Br_MaxEnt_range_compare_true} shows the estimated posterior mean model does contain useful information on CMB field structures above spherical harmonic degree 13.  Field structures remain coherent as power is added from degree 13 to 22, with some important details recovered.  For example, in the southern polar region, under the Australian-Antarctic basin (near latitude 60$^\circ$S, longitude 135$^\circ$E) there is a localized intense flux feature present in both the dynamo model and in the posterior mean; this is weaker and smeared when the CMB field is truncated at degree 13.  Similarly in the northern polar region under Siberia and Alaska, both intense normal polarity features and reversed flux features are better retrieved in the posterior mean model to degree 22 compared to a more conventional model truncated at degree 13.  

At low latitudes a number of intense features are better retrieved in the estimated posterior mean model to degree 22 than in the model truncated at degree 13, for example the flux concentration under central Africa and the positive-negative pair of flux patches arranged north-south across the equator under India and under the equator south of Mexico.  However some smaller scale features, for example a positive flux concentration in the dynamo model under the equator at 15 degrees west, are still poorly retrieved.  Isolated small scale flux features with no imprint on larger scales, and which change rapidly in time, are not well recovered.

Flux features recovered in the posterior mean core field model up to degree 22 are however always related to features in the true dynamo field. We do not find any evidence for artefacts due to leakage of the lithospheric field.  The reconstructed fields are by construction as simple as possible (in terms of maximizing the information entropy) while satisfying the observational constraints.  The adopted temporal prior (spline smoothing) also involves strong time-averaging over small length scales that is absent in the dynamo, this will contribute to the loss of detail on small scales.   We note that interpretations in terms of the model resolution matrix \citep[e.g.][]{Bloxham_Gubbins_Jackson_1989} are not straightforward here.  With the available satellite data coverage we are able to well resolve the internal field up to high spherical harmonic degree. It is however constraints from prior information that allow us to partially separate the core and lithospheric fields. 

Variations are found in the synthetic test results regarding the amount of small scale power in the posterior mean model as the data constraints change. For example there is less power in the small scale field in the gap between the CHAMP and \textit{Swarm} missions when only high altitude and lower quality satellite data from Cryosat-2 was available.  This is an expected consequence of the maximum entropy method \-- at times when the data constraints are weaker the posterior mean model becomes simpler and the model uncertainty at high degree larger.

\subsection{Interpretations of the inferred CMB field}

\begin{figure}
	\centerline{\includegraphics[width=\textwidth]{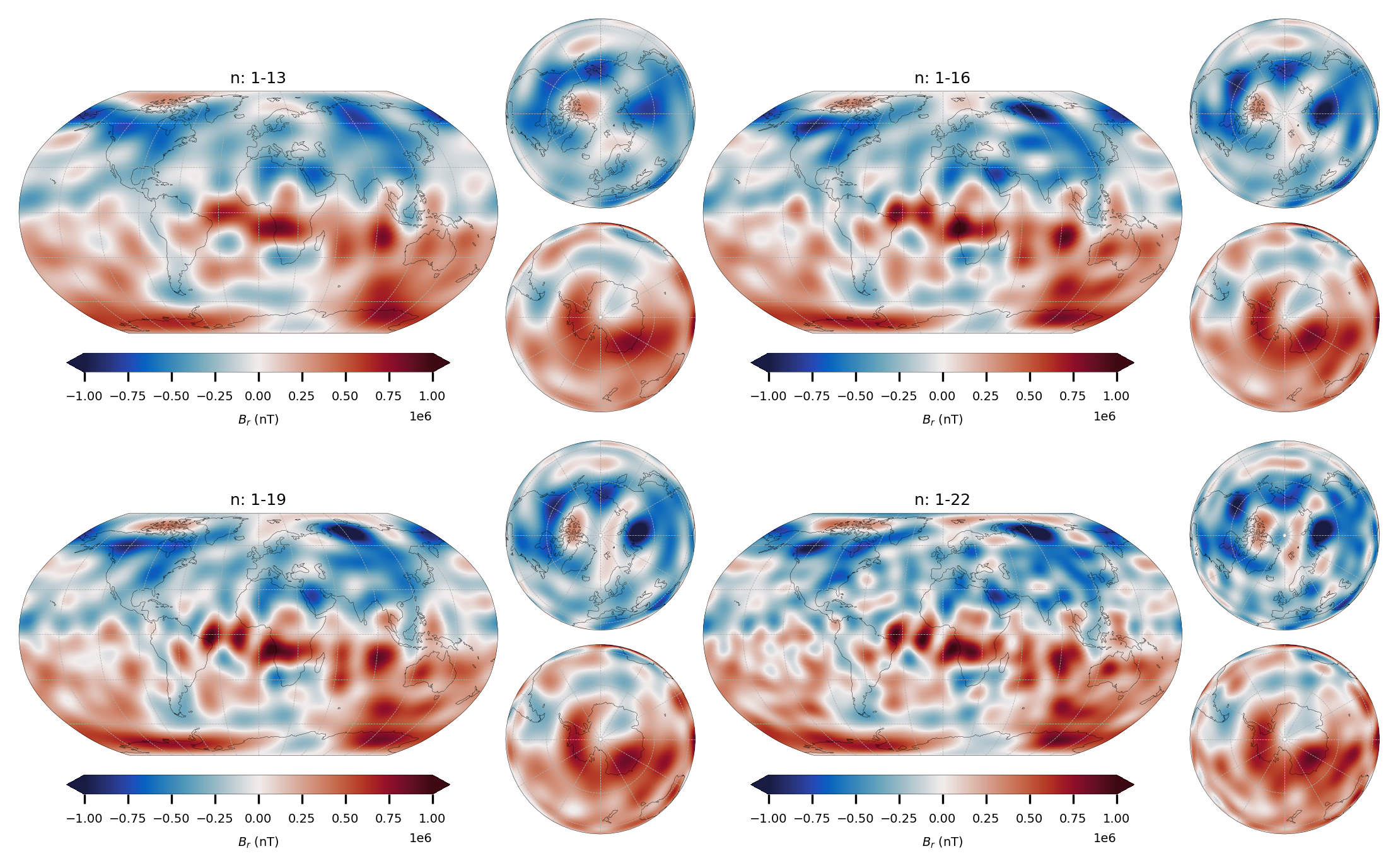}}
	\caption{Maps of radial magnetic field at the core-mantle boundary in 2020.0 from the model CL posterior mean, as a function of the truncation degree, up to degree 13 (top left), to degree 16 (top right), to degree 19 (bottom left) and to degree 22 (bottom right).  Note how the power adds coherently to existing features already evident at lower truncation degrees.}
	\label{fig:Br_MaxEnt_core_nmax_range}
\end{figure}

Returning to the model CL derived from the real observations, in Fig.\,\ref{fig:Br_MaxEnt_core_nmax_range} we present maps of the radial magnetic field at the CMB from the posterior mean model, for increasing truncation degrees of 13, 16, 19 and 22.  As in the synthetic test, power adds coherently with increasing degree with some structures becoming more localized and intense. This is also the case for the SV at the CMB, which we have examined in manner similar to \citet{Holme:2011} (not shown); SV features seen at lower degree remain coherent although more small scale SV features begin to appear by degree 22 indicating this is at the edge of what we are able to reliably study.  

At high latitudes strong normal flux features are found close to the where the inner core tangent cylinder intersects the CMB (at latitudes 69.6 degrees north and south). There is a particularly intense flux concentration under the Taymyr Peninsula in Siberia in 2020, and a number of high amplitude features arranged near latitude 60 degrees north under Greenland, Canada and eastern Siberia.  In the Southern polar region there are strong normal flux features under Wilkes land in eastern Antarctica and under western Antartica near the Antarctic peninsula. Normal flux patches localized close to the tangent cylinder are consistent with the poloidal dipole field being produced by an alpha-effect in energetic eddies originating in vigorous convection close to the inner core boundary. Another possibility is that such eddies are spawned by powerful azimuthal flows inside the tangent cylinder occasionally ejected across the tangent cylinder \citep[e.g.][]{schaeffer2017turbulent}.  The different locations of the flux concentrations in the northern and southern hemisphere seem difficult to explain in terms of purely columnar flows. 

Reversed flux features are found in the northern polar region under the Canadian Arctic, centred under the Queen Elizabeth Islands, and also under the Nansen Basin (between Spitzbergen and the new Siberian Islands).  In the Southern polar region there is an extended but weak reverse flux feature under Eastern Antarctica southwards from Africa, this feature, also present in models truncated at degree 13, persists to higher degree and is interesting as it crosses the tangent cylinder. Reverse flux features inside the tangent cylinder could be related to a strong omega effect driven by strong azimuthal flows inside the tangent cylinder and associated flux expulsion, similar features have been seen in turbulent dynamos \citep{schaeffer2017turbulent, Sheyko_2018}. 

At low latitudes the strong flux feature found under the equator between Africa and south America in models truncated at degree 13 is split into two features, as is a normal feature under central Africa.  Such splitting of low latitude flux features has been suggested in previous attempts to retrieve the core field above degree 13 \citep{Baerenzung_2020, Otzen:2022b}, and it is consistent with the patterns of core surface SV retrieved above degree 13 \citep{Finlay_CHAOS7:2020}.  Many of the field concentrations at low latitudes seem to occur in oppositely signed pairs and they are observed to move westwards.  A possible explanation could be that they are the signature of toroidal flux being expelled from the core at low latitudes and subsequently propagating as a wave \citep[e.g.][]{Aubert:2013, Aubert:2022}.  Such features are ubiquitous at low latitudes in strongly forced geodynamo simulations when the viscosity is sufficiently low \citep{Sheyko_2014,schaeffer2017turbulent, Aubert2019}.

\subsection{Features of the inferred large-scale lithospheric field}
\begin{figure}
	\centerline{\includegraphics[width=0.8\textwidth]{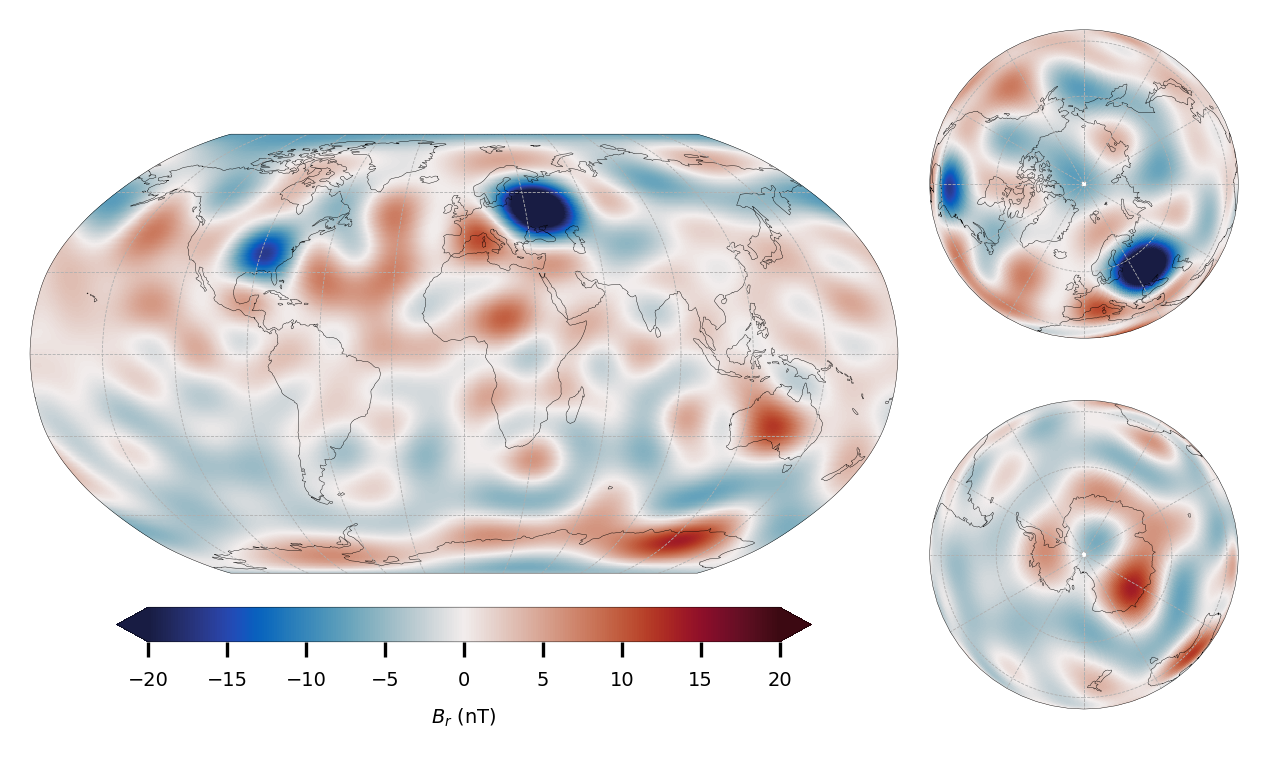}}
	\caption{Estimated posterior mean large-scale lithospheric field (degrees 2 to 14) plotted on a spherical surface at Earth's mean reference radius.}
	\label{fig:Br_MaxEnt_lith_low_deg}
\end{figure}

In the synthetic test reported in Appendix \ref{sec:AppA} we were also able to test the retrieval of the lithospheric field.  The lithospheric field at the Earth’s surface generally compares well with the input synthetic truth lithospheric field, albeit with less power below degree 15 and above degree 90 (Fig.\ref{fig:pspec_spatial_synth} and Fig.\,\ref{fig:Br_MaxEnt_lith_synth}).  Of particular interest is whether or not any details of the synthetic truth large scale lithospheric field were retrieved.  Fig.\,\ref{fig:Br_MaxEnt_lith_synth_large} shows that, somewhat surprisingly,  some details of the large scale lithospheric field can be recovered, albeit with reduced amplitude.  The largest anomalies in the recovered large scale lithospheric field, for example between the North Pole and the Bering strait, near Australia and in north-Eastern Europe are also present in the synthetic truth model. On the other hand some prominent structures are missing or incomplete, for example in the  Atlantic-Indian-Antarctic basin or under North America, and the recovered amplitudes are too low.  Fig. \ref{fig:Br_MaxEnt_lith_low_deg} shows a similar map of the large scale lithospheric field from the model CL posterior mean derived from the real data.  It shows strong anomalies in the northern part of Eastern Europe, Australia, around eastern Antarctica and under eastern North America, which are known locations of strong continental magnetic anomalies.  

\subsection{Limitations and future prospects}
The aim of this study was to better separate core and lithospheric magnetic fields; it has only been partially successful.  The recovered small scale core field and large scale lithospheric field lack power. Use of more informative priors, if these can be justified, would certainly bring improvements.  For example, we used a single spatial covariance function for each source,  which assumes field structures that are statistically the same for all locations on the sphere. The ensembles of prior fields from the dynamo and magnetisation models could instead be used to build full covariance matrices characterising the statistical covariances between all locations on the sphere for each source.  Such dense spatial covariance matrices have already been used by other authors \citep{Gillet:2019,Ropp_2020,Istas:2023}, in the context of lower resolution core field and flow modelling.  Once sufficiently large prior ensembles are available this will be a relatively simple extension of the method presented above.  A concern with this approach is that imperfect aspects of the dynamo and magnetisation simulations might be mapped into the estimated field models, it was for this reason we started here by using rather simple information from the prior ensembles.  

Regarding the small scale core field, the spline-based temporal prior employed here prevents the recovery of rapid changes on small length scales. This limitation could be particularly serious at low latitudes.  This situation could be remedied by adopting temporal priors that better reflect the expected physics, for example based on AR2 or AR3 processes \citep{Gillet2013,Sadhasivan_Constable_2022}, or perhaps using temporal statistics from high resolution simulations \citep{Aubert:2023}.  

A variant of the approach presented here is to take the spherical harmonic (Gauss) coefficients of the internal potential as input data for the separation into core and lithospheric fields rather than satellite data.  This has some computational advantages and may prove useful in future applications, further details and an example are presented in Appendix \ref{sec:AppB}.  Our scheme could also be applied to dedicated studies of the lithospheric field.  This would require improving the lithospheric prior to allow for more power at small length scales, use of higher data sampling rates, and use of \textit{Swarm} east-west gradients that were not included in this study.

Much is still to be learnt regarding the small scale core field.  The maps of the posterior mean core field presented here show how information on the core dynamo is lost when CMB fields are truncated at degree 13.    On the other hand our ability to retrieve the small scale core field, and avoid lithospheric field contamination, depends on correctly formulating and utilizing prior information regarding the sources.  Further effort is needed on how best to extract reliable prior information from a variety of simulations of the core dynamo and the lithospheric magnetisation.  Improved Bayesian field modelling requires prior ensembles that are both informative and broadly representative of the diversity of possible core and lithospheric fields.

% Better than hard truncation at degree 13.  

%Not fully satisying as posterior spread larger than mean values, at least in the spectrum.

%Provides framework for furture.

% other grids with better conditioning

% posterior errors depend on spread of prior, could be chosen better

%===============================================================================
\section*{Availability of datasets and material}
The data used in this article and the derived field models are archived at \url{https://doi.org/10.11583/DTU.24968763.v1}. Magnetic field data from the \textit{Swarm} mission are freely available from \url{https://earth.esa.int/web/guest/swarm/data-access}; CHAMP data are available from \url{https://isdc.gfz-potsdam.de/champ-isdc}; Ground observatory data are available from \url{ftp://ftp.nerc-murchison.ac.uk/geomag/Swarm/AUX_OBS/hour/}; The RC-index is available from \url{http://www.spacecenter.dk/files/magnetic-models/RC/}; The CHAOS-7 model and its updates are available at \url{http://www.spacecenter.dk/files/magnetic-models/CHAOS-7/}; solar wind speed, interplanetary magnetic field, and Kp-index are available from \url{https ://omniw eb.gsfc.nasa.gov/ow.html}.
%===============================================================================
\begin{acknowledgments}
The authors thank Julien Aubert for providing data from his dynamo simulations and Simon Williams for help with the lithospheric magnetisation simulations.  Nils Olsen and Lars T{\o}ffner Clausen are thanked for numerous helpful discussions. Andy Jackson is thanked for discussions regarding maximum entropy and correlation functions. Nicolas Gillet and Phil Livermore provided constructive comments on an earlier version of this work that appeared in MO's PhD thesis.  The editor Richard Holme and two anonymous reviewers are thanked for comments that helped us improve the manuscript. We thank the GFZ German Research Centre for Geoscience for providing access to the CHAMP MAG-L3 data and the European Space Agency (ESA) for providing prompt access to the {\it Swarm} L1b data. The high resolution 1-min OMNI data were provided by the Space Physics Data Facility (SPDF), NASA Goddard Space Flight Centre. We thank the staff of the geomagnetic observatories and the INTERMAGNET for providing high-quality observatory data. This work was funded by the European Research Council (ERC) under the European Union’s Horizon 2020 research and innovation programme (grant agreement No. 772561).  
\end{acknowledgments}
%===============================================================================
\bibliographystyle{gji}
\bibliography{refs}
%\nocite{*}
%=============================================================================== 
%\newpage
%=============================================================================== 

\appendix
\section{An application to synthetic data}
\label{sec:AppA}

Here we report results from a synthetic test designed to test the extent to which we can retrieve the small scale core field above spherical harmonic degree 13 and the large scale lithospheric field below degree 13 using the scheme described in Section \ref{sec:method}.

We use as input a time-dependent core field taken from a preliminary version of the dynamo model assimilation runs described by \citep{Aubert:2023}.  This simulation contained realistic core field structures and time dependence up to degree 30 and was not contained in the prior ensemble.  For the synthetic lithospheric field we used one realizatation from our set of prior magnetisation models, based on the forward models by \citet{Hemant_Maus_2005}, \citet{Masterton_2013} and \citet{Williams_2019} with the perturbations described by \citet{Otzen_2022}, but generated separately and not included in the ensemble used to construct the prior statistics.  From these models we synthesized data at the same times and positions, and with the same measured components, as the real satellite and ground data described in Section \ref{sec:data}. Gaussian noise was added using a standard deviation of $3.0$nT for CHAMP, $5.0$nT for CryoSat-2, $2.0$nT for \textit{Swarm} and $3.0$nT/yr for ground observatory data. 

Inversions were then carried out in exactly the same way as for the real data. The same priors, and same fixed regularization settings the same starting models and the same number of iterations (24) were performed.  

\begin{figure}
	\centering\includegraphics[width=0.7\textwidth]{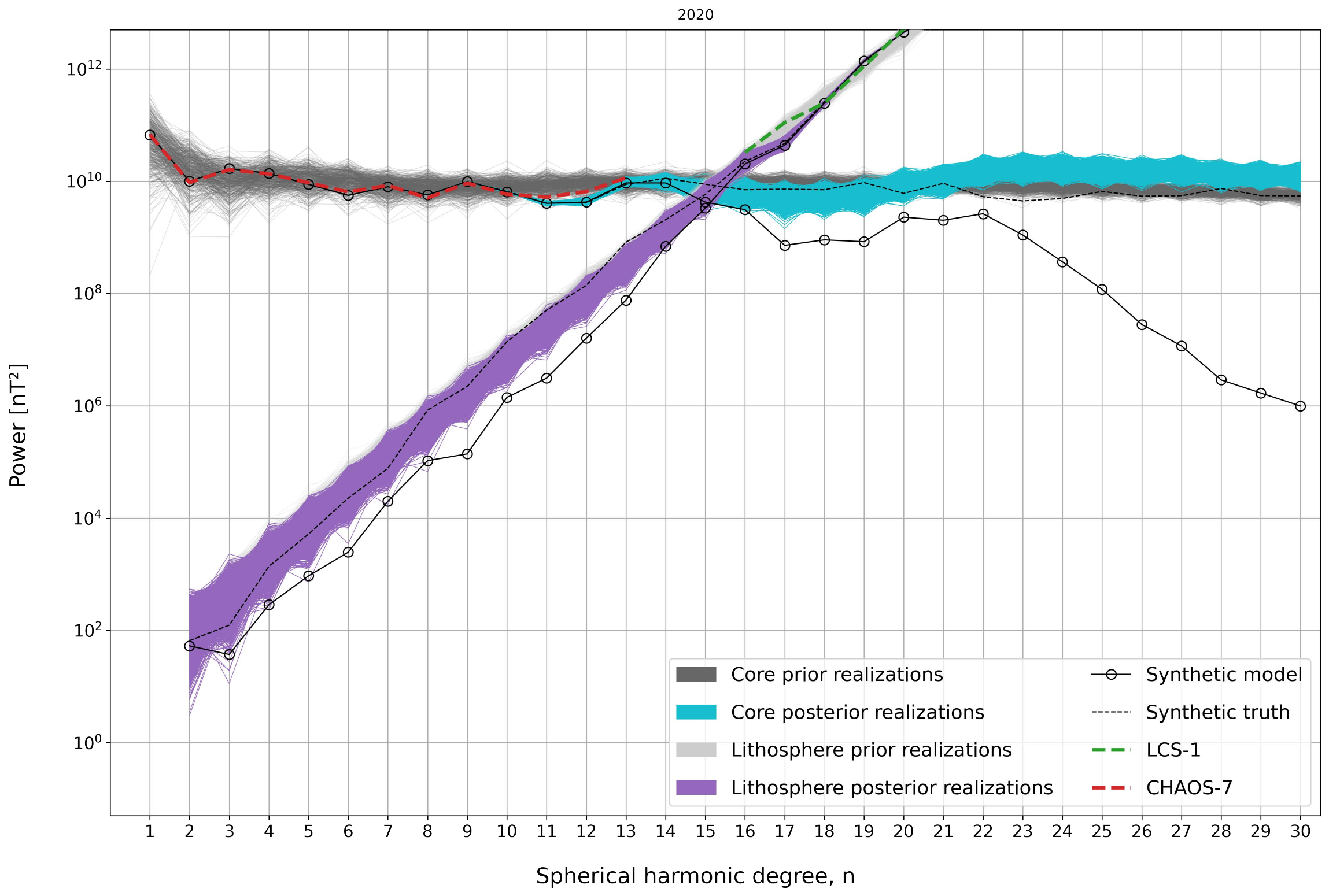}\\
 \centering\includegraphics[width=0.7\textwidth]{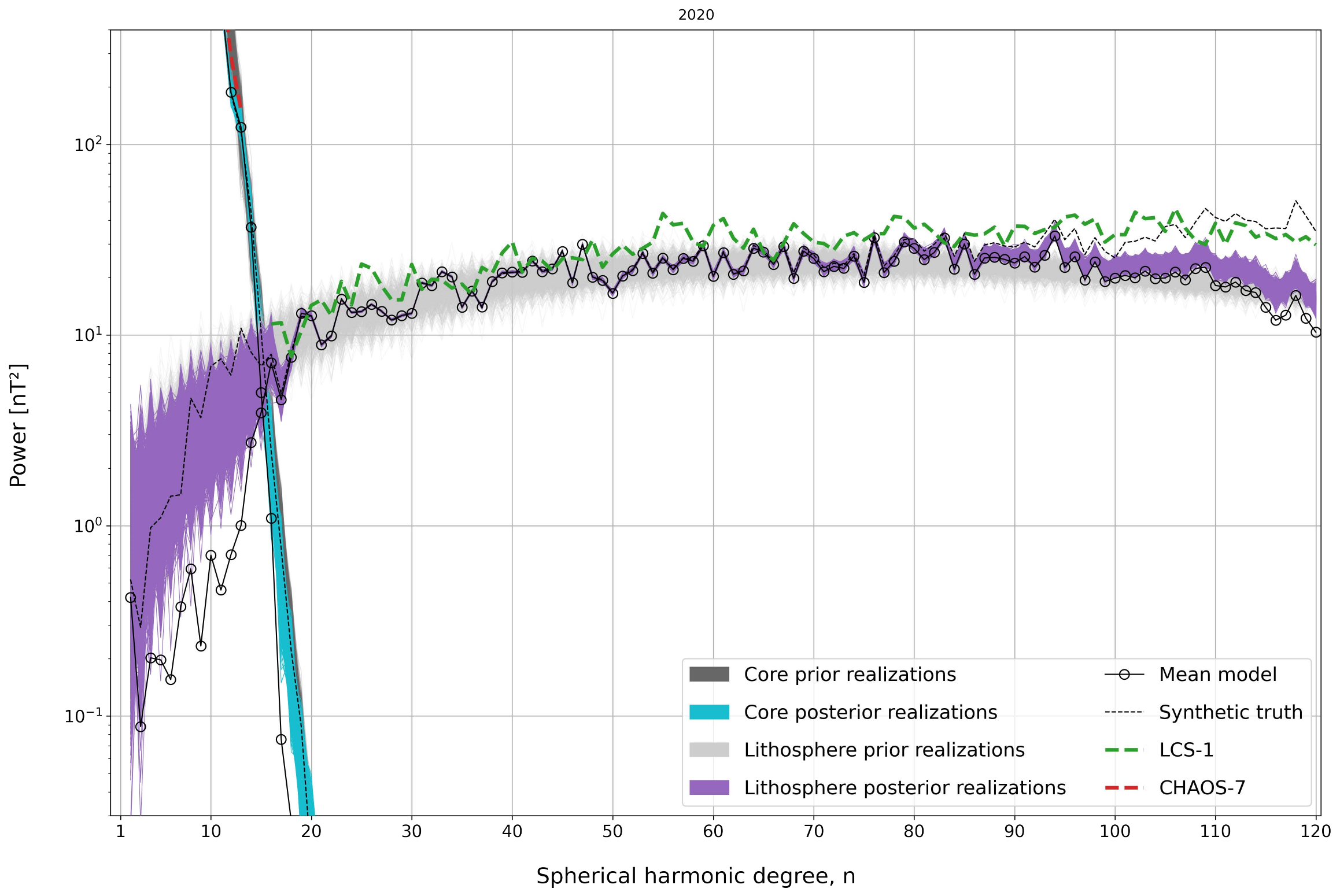}
    
	\caption{Lowes-Mauersberger spectra showing power as a function of spherical harmonic degree of magnetic fields at the CMB (top) and at Earth's surface spherical reference radius, (bottom) in 2020.0. The input synthetic truth models for the core and lithospheric fields are shown in the black dashed lines.  Solid black lines with circles show the estimated posterior mean core and lithospheric field models which cross at degree 15. Grey lines show prior realizations based on the adopted spatial covariance model, cyan lines show posterior realizations of the core field and purple lines posterior realizations of the lithospheric field.  CHAOS-7 (up to degree 13) and LCS-1 (at degree 16 and above) are shown for reference.}
	\label{fig:pspec_spatial_synth}
\end{figure}

\begin{figure}
	\centerline{\includegraphics[width=\textwidth]{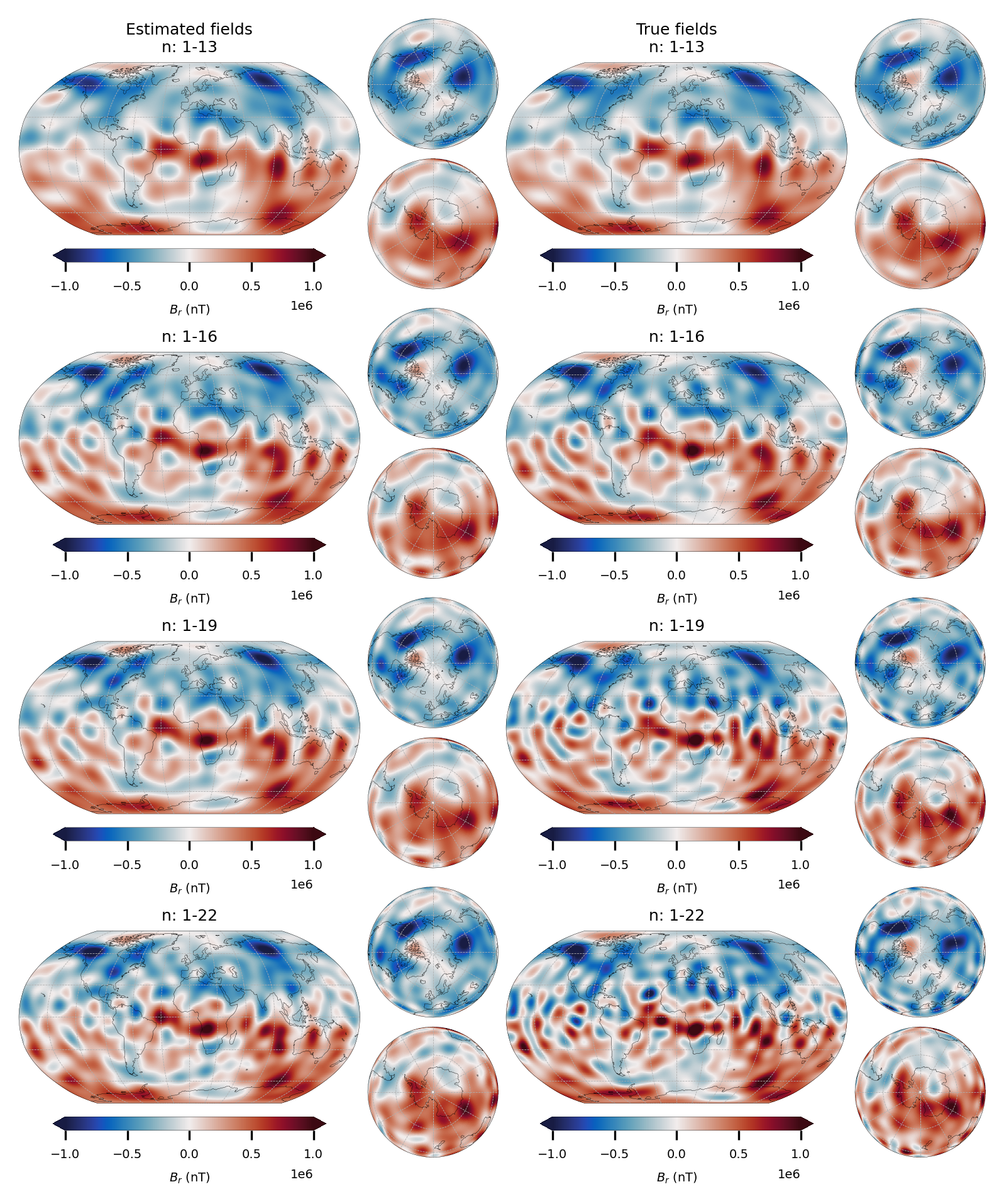}}
	\caption{Maps of radial magnetic field at the core-mantle boundary from the estimated posterior mean model derived from the synthetic test dataset (left column) and the input truth dynamo model (right column), as a function of the truncation degree, up to degree 13 (top row), to degree 16 (second row), to degree 19 (third row) and to degree 22 (bottom row).}
	\label{fig:Br_MaxEnt_range_compare_true}
\end{figure}

In Fig.\, \ref{fig:pspec_spatial_synth} we present the Lowes-Mauersberger spectrum at the core surface (top) and at the Earth's surface (bottom), with the synthetic truth marked in the dashed line.  In Fig.\, \ref{fig:Br_MaxEnt_range_compare_true} we present how the estimated posterior mean and the synthetic truth for the core field changes with the truncation degree.  Although some small scale feature are lost and the amplitude is reduced, the posterior mean model above degree 22 retrieves more details of real features.

%\begin{figure}
%	\centerline{\includegraphics[width=\textwidth]{./FIGS/appendix/maps/Br_MaxEnt_core_synth_compare_dynamo.png}}
%	\caption{}
%	\label{fig:Br_MaxEnt_core_synth}
%\end{figure}

\begin{figure}
	\centerline{\includegraphics[width=0.9\textwidth]{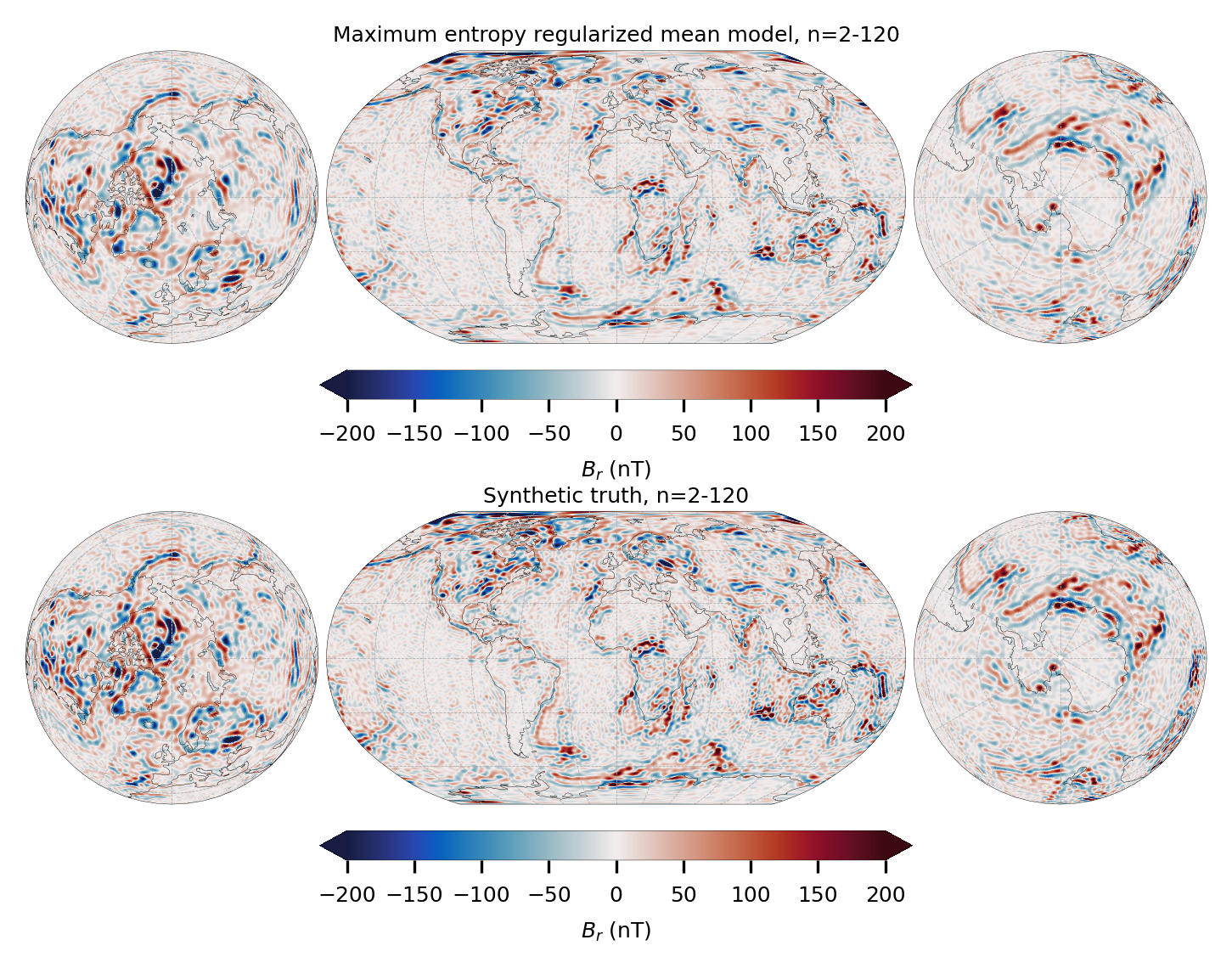}}
	\caption{Map of the radial component of the estimated posterior mean lithospheric field on a spherical surface at Earth's reference radius, for degrees 2 to 120 (top) and the same quantity from the synthetic truth input model based on lithospheric magnetisation models (bottom).}
	\label{fig:Br_MaxEnt_lith_synth}
\end{figure}

\begin{figure} 
 \centerline{\includegraphics[width=0.9\textwidth]{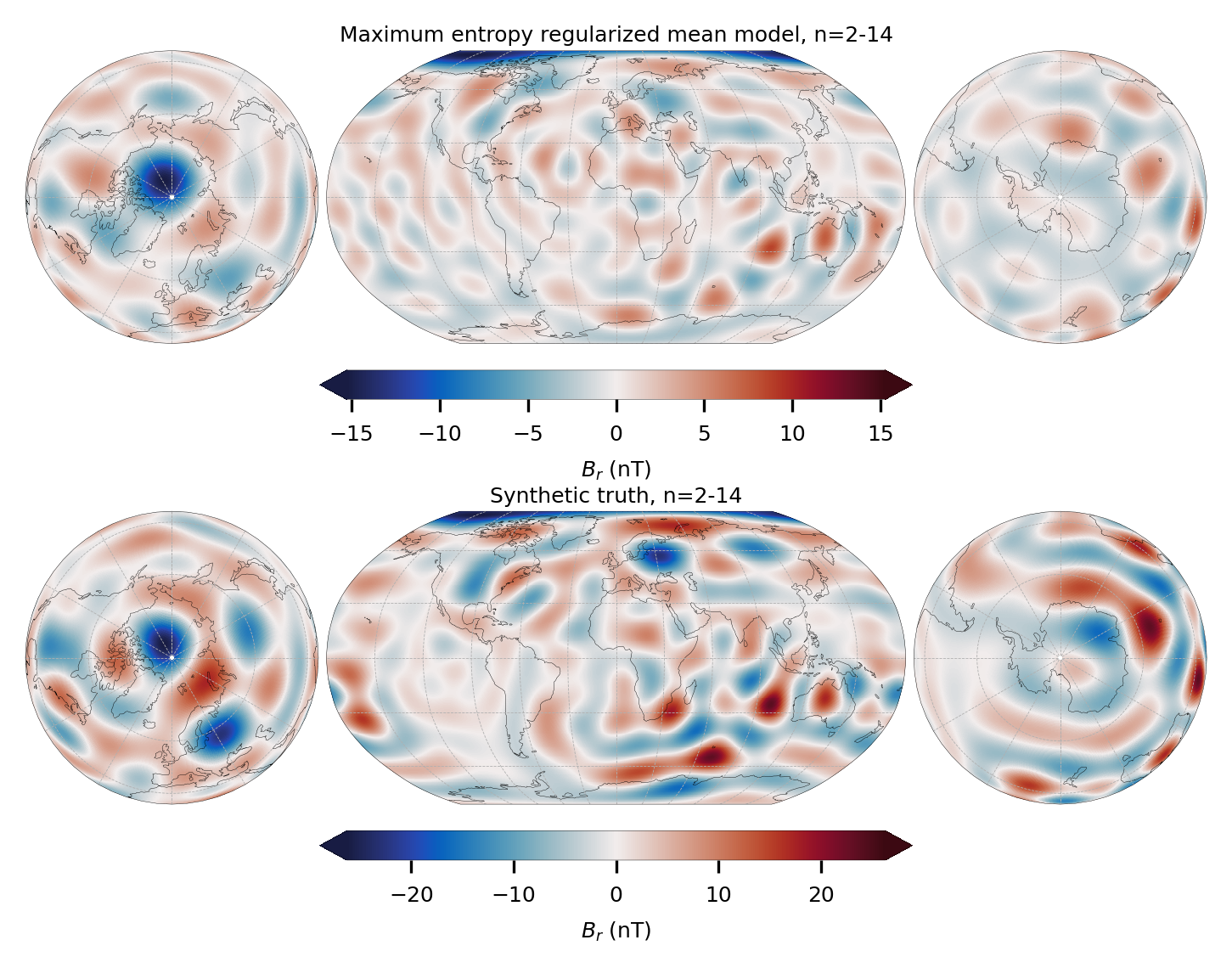}}
	\caption{Estimated posterior mean large-scale lithospheric field  (degrees 2 to 14) derived from the synthetic dataset plotted at Earth's spherical reference surface (top) and the same quantity from the synthetic truth model based on a prior realization of lithospheric magnetisation (bottom).  Note the change in the scale used for the two panels. }
	\label{fig:Br_MaxEnt_lith_synth_large}
\end{figure}

Fig.\, \ref{fig:Br_MaxEnt_lith_synth} presents a comparison of the estimated posterior mean lithospheric field and Fig.\,\ref{fig:Br_MaxEnt_lith_synth_large} compares the recovered (posterior mean) large scale lithospheric field, and the input synthetic truth for the large scale lithospheric field at Earth's spherical reference radius, for degrees 2-14.

\clearpage

\section{Application to Gauss coefficients of the internal field}
\label{sec:AppB}
Instead of co-estimating core and lithospheric fields models directly from satellite observations it is possible to start with Gauss coefficients for an internal potential field, $g^{int}_{nm}(t_i)$ and $,h^{int}_{nm}(t_i)$, given at a series of reference times $t_i$.  In this case the relevant input data consisting of the Gauss coefficients at all times can be collected in a vector 
\begin{equation}
\mathbf{g}=\left[g^{int}_{10}(t_0), g^{int}_{11}(t_0), h^{int}_{11}(t_0)..., g^{int}_{10}(t_1), ... \right]^T.
\end{equation}
If the input field model is already smooth in time, no additional temporal prior is needed and the loss function to be minimized takes the form   
\begin{equation}
 \Phi(\mathbf{m}) = \frac{1}{2}\chi_g^{2}(\mathbf{m}) -  S_{tav} (\mathbf{x}^{C}) - S (\mathbf{x}^{L}). 
\end{equation}
where $\chi_g^{2}=\mathbf{e_g}^T \mathbf{C_g^{-1}} \mathbf{e_g}$ with $\mathbf{e_g} = \mathbf{g} - \widehat{\mathbf{g}}$ and $\widehat{\mathbf{g}}$ are the predicted internal field Gauss coefficients (core plus lithosphere) at the relevant times and the diagonal values of $\mathbf{C_g^{-1}}$ are $1/\sigma_g^2$ which define how closely the input Gauss coefficients should be matched. In the experiments reported here we set $\sigma_g=10^{-5}$nT since we wanted the estimated core and lithospheric coefficients to closely match those of the input field model.  As before $S_{tav}(\mathbf{x}^{C})$ describes the information entropy of the spatially decorrelated CMB radial field (averaged over time), and $S (\mathbf{x}^{L})$ describes the information entropy of the spatially decorrelated lithospheric radial field at Earth's surface.

\begin{figure}
    (a) \qquad  \qquad \qquad \qquad \qquad \qquad \qquad  \qquad \qquad \qquad  (b)\\
          \includegraphics[width=0.49\textwidth]{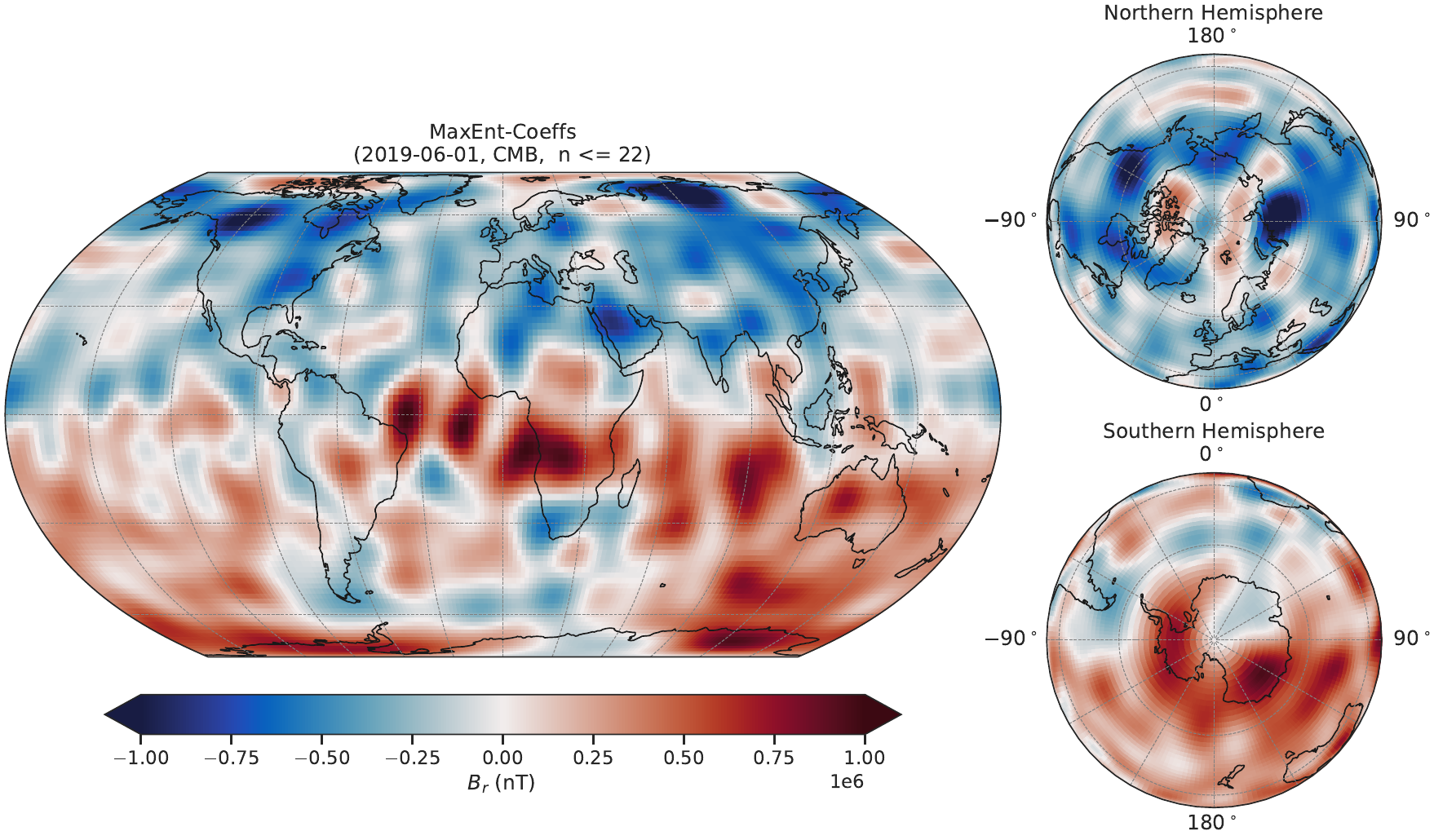}
    \includegraphics[width=0.49\textwidth]{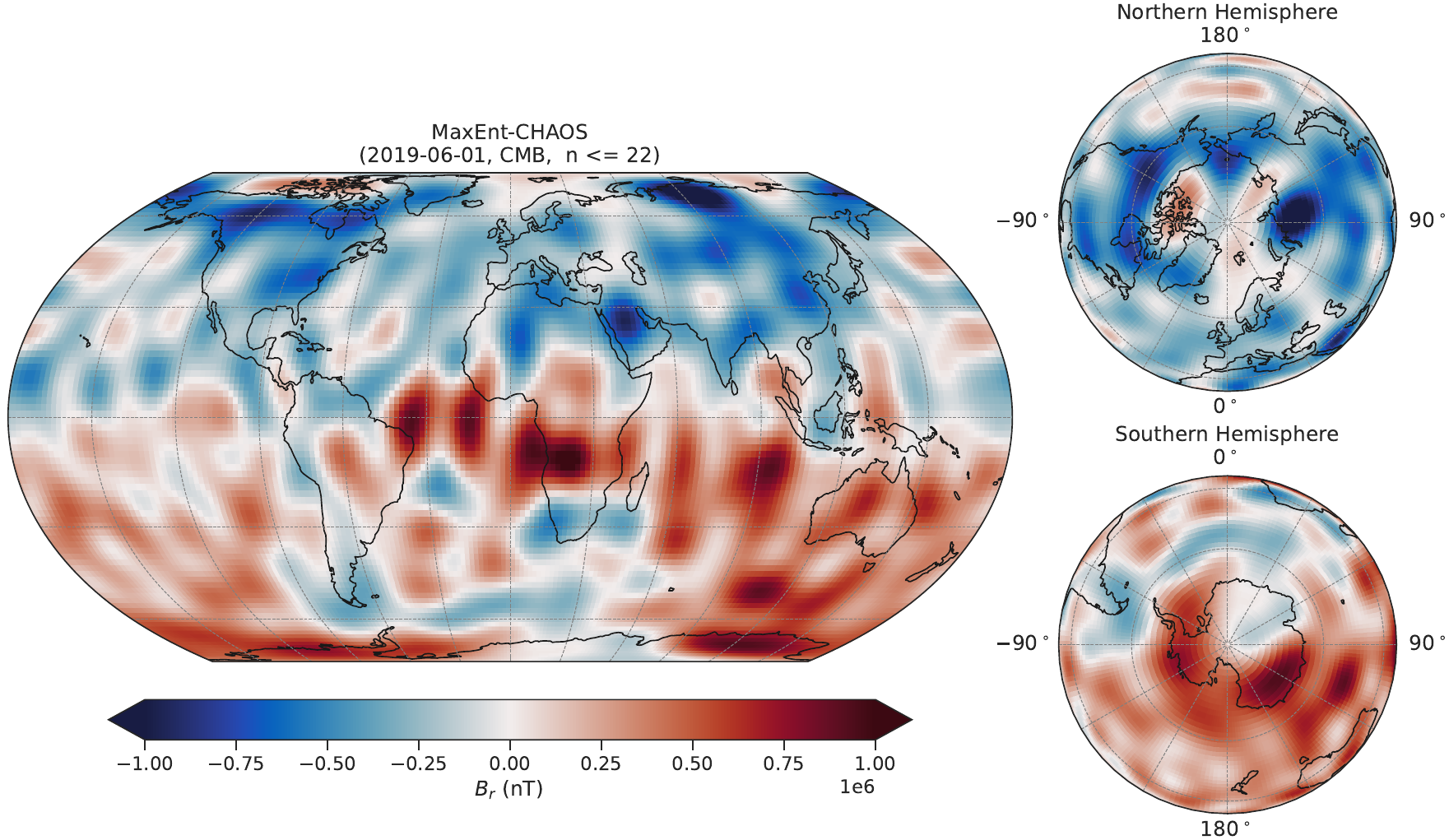}\\
     (c) \qquad  \qquad \qquad \qquad \qquad \qquad \qquad  \qquad \qquad \qquad  (d)\\
    \includegraphics[width=0.49\textwidth]{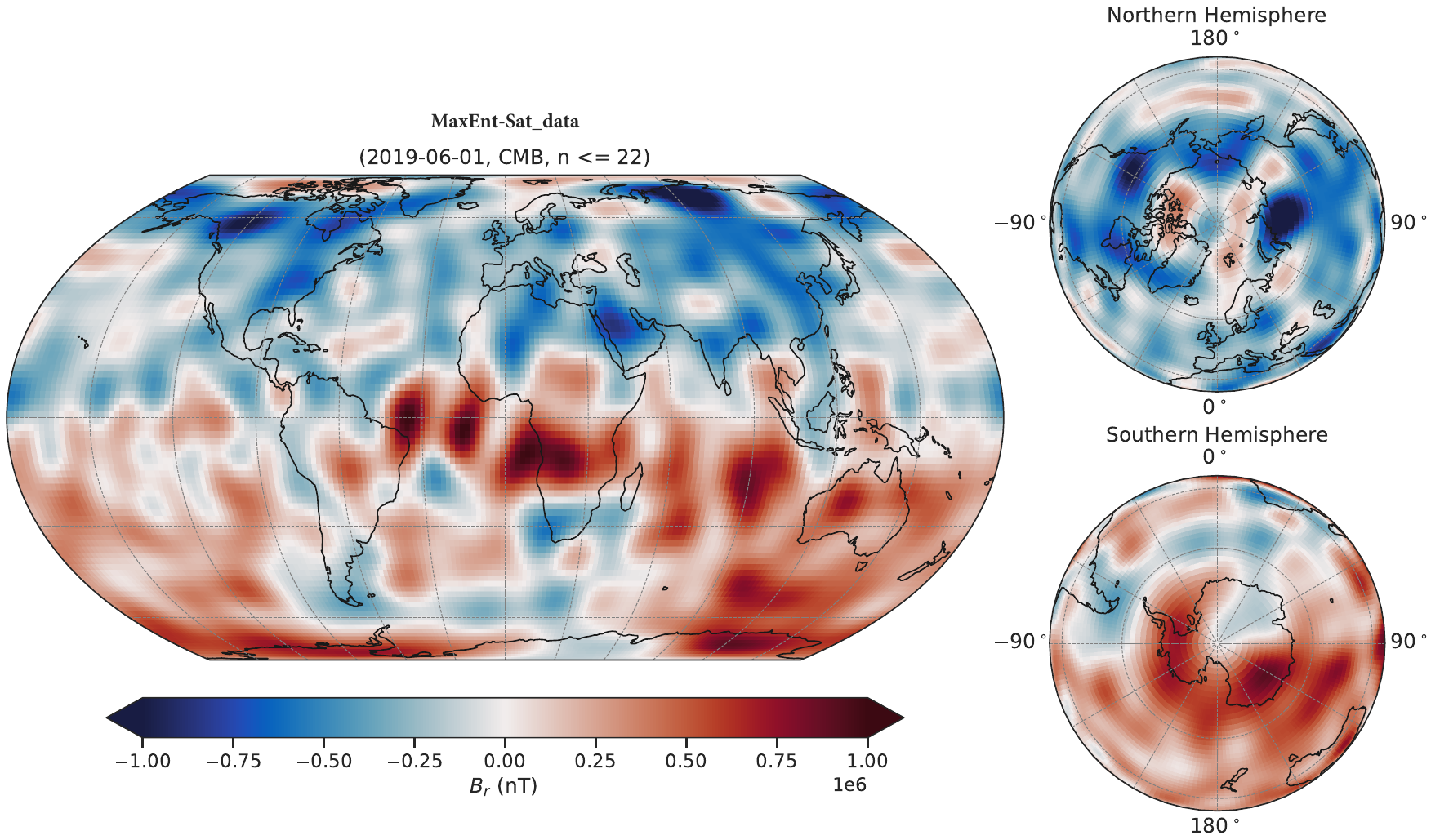} \includegraphics[width=0.49\textwidth]{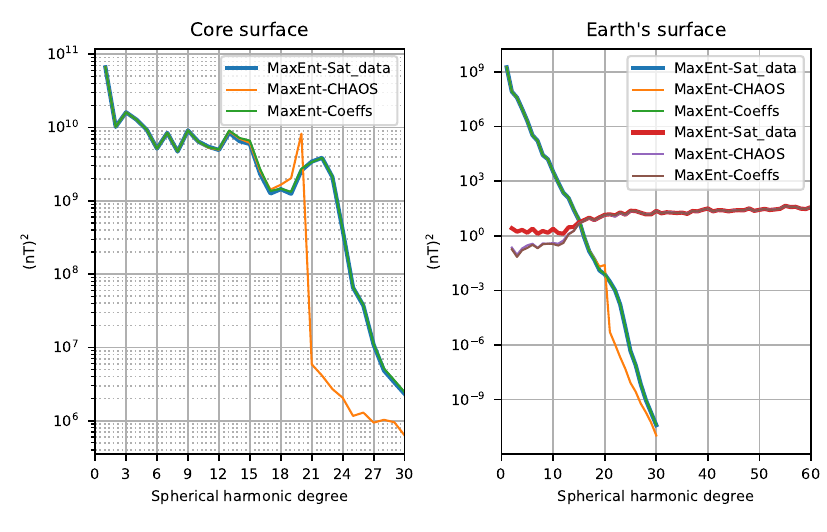}\\
    \caption{ (a) Radial magnetic field at the core-mantle boundary in 2019.5, from a model estimated taking as the input combined internal field spherical harmonic coefficients (core and lithospheric fields) from the CL model presented in the main text (top left, denoted MaxEnt-Coeffs), (b) from a model estimated in a similar way but taking spherical harmonic coefficients from CHAOS-7.16 as input data (top right, denoted MaxEnt-CHAOS), (c) for reference, the result from the CL model presented in the manuscript (bottom left, denoted MaxEnt-Sat$\_$data), (d) Spherical harmonic power spectra for the three models at the CMB and at Earth's surface (bottom right).}
    \label{Fig:MaxEnt_coeffs}
\end{figure}

Results of applying this procedure to the total internal field (core plus lithosphere) from the CL model from the main text, and to the CHAOS-7.16 internal field model are shown in Fig. \ref{Fig:MaxEnt_coeffs}. In these tests we started from an initial lithospheric field set to zero below degree 16 and to the LCS-1 model at higher degree and an initial core field set to the time-average of the time-dependent internal field from the CHAOS-6.9 model up to degree 12 and zero at higher degree, and we found convergence after 5 to 6 iterations. 

We find the CMB field retrieved in this way is rather similar starting with CHAOS-7.16 and the full CL model, with the minor differences in 2019.5 seen in Fig. \ref{Fig:MaxEnt_coeffs} being due to (i) the limitation of the CHAOS-7.16 time-dependent field to degree 20 and possible aliasing related to this, and (ii) because CHAOS-7.16 is derived using data up to 2023 while the CL model used data only up to 2020.  On the other hand we find that the large-scale lithospheric fields estimated here from input Gauss coefficients contain slightly lower power than that estimated from satellite observations, perhaps because there is less freedom here when fitting the input data.  We find that time-dependence of the input internal field is crucial to the separation.

\begin{figure}
    \centering
    \includegraphics[width=0.8\textwidth]{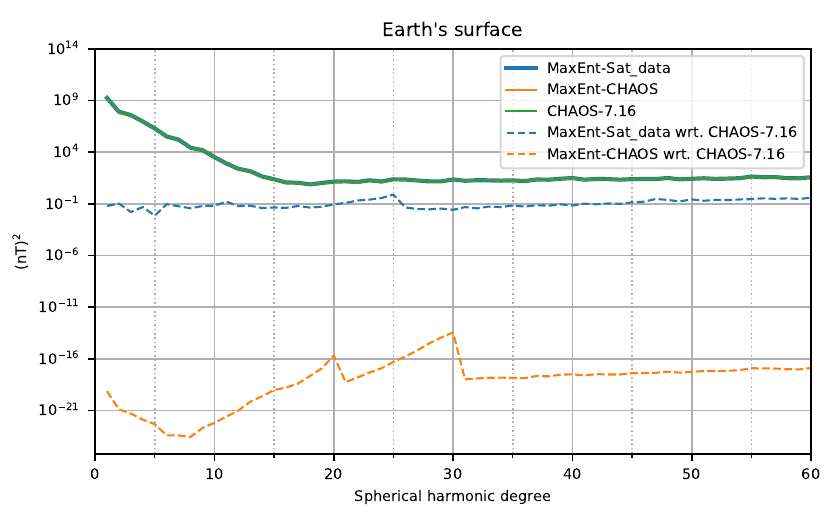}
    \caption{Comparison of spatial power spectra of the combined (core plus lithosphere) internal field  for the models CHAOS-7.16 (green), the CL model from the main text labelled MaxEnt-Sat$\_$data (blue) and the model from this Appendix, labelled  MaxEnt-CHAOS, derived from the spherical harmonic coefficients of CHAOS-7.16.  Difference to CHAOS-7.16 are marked as dashed lines.}
    \label{Fig:Compare_combined_spectra}
\end{figure}

The power spectra of the combined internal field from these estimated models is presented and compared to that of CHAOS-7.16 in Fig. \ref{Fig:Compare_combined_spectra}.  The spectra of the combined internal field model derived from the CHAOS model coefficients (labelled MaxEnt-CHAOS) matches the CHAOS spectra to machine precision \-- this was enforced by construction.  The CL model from the main text (labelled MaxEnt-Sat$\_$data) is, as expected, not identical to CHAOS, but it nevertheless agrees well at all degrees, the difference being less than 0.1\,nT$^2$ in the power (i.e. around 0.3\,nT) which is less than the likely errors in the model coefficients.  The combined CL model is thus compatible with the observational constraints, in that it fits the satellite data from which it was constructed, and its combined internal field agrees well with established field models such as the CHAOS model.

The scheme outlined in this Appendix provides a means of separating a given internal field model into core and lithospheric parts as a post-processing procedure; this may prove useful for future applications.
 
\end{document}